\documentclass[twocolumn,tighten,numberedappendix,appendixfloats]{aastex6}
\usepackage{natbib}
\usepackage{units}
\usepackage{amsfonts}
\usepackage{amsmath}

\newcommand{\overbar}[1]{\mkern 1.5mu\overline{\mkern-1.5mu#1\mkern-1.5mu}\mkern 1.5mu}

\newcommand\litl{\rm\scriptscriptstyle}

\newcommand{\mum}{\ensuremath{\,\mu{\mathrm m}\/}}
\newcommand{\mjysr}{\ensuremath{{\,\rm MJy}\, {\rm sr}^{-1}}}
\newcommand{\Spitzer}{\hbox{\sl Spitzer\/}}
\newcommand{\ppm}[2]{\ensuremath{^{+{#1}}_{-{#2}}}}

\slugcomment{Astronomical Journal, Accepted 2016 May 26}

\shorttitle{Repeatability of Spitzer Eclipse Depths}
\shortauthors{Ingalls et al.}

\begin{document}


\title{Repeatability and Accuracy of Exoplanet Eclipse Depths Measured with Post-Cryogenic {\sl Spitzer}}


\author{James G. Ingalls\altaffilmark{1}, J. E. Krick\altaffilmark{1}, S. J. Carey\altaffilmark{1}, John R. Stauffer\altaffilmark{1}, Patrick J. Lowrance\altaffilmark{1}, Carl J. Grillmair\altaffilmark{1}, Derek Buzasi\altaffilmark{2}, Drake Deming\altaffilmark{3}, Hannah Diamond-Lowe\altaffilmark{4}, Thomas M. Evans\altaffilmark{5}, G. Morello\altaffilmark{6}, Kevin B. Stevenson\altaffilmark{4}, Ian Wong\altaffilmark{7}, Peter Capak\altaffilmark{1}, William Glaccum\altaffilmark{1}, Seppo Laine\altaffilmark{1}, Jason Surace\altaffilmark{1}, Lisa Storrie-Lombardi\altaffilmark{1}}





\altaffiltext{1}{{\sl Spitzer} Science Center, California Institute of Technology, 1200 E California Blvd, Mail Code 314-6, Pasadena, CA 91125, USA; ingalls@ipac.caltech.edu}
\altaffiltext{2}{Department of Chemistry and Physics, Florida Gulf Coast University, Fort Myers, FL 33965, USA}
\altaffiltext{3}{Department of Astronomy, University of Maryland, College Park, MD 20742-2421, USA}
\altaffiltext{4}{Department of Astronomy and Astrophysics, University of Chicago, 5640 S Ellis Avenue, Chicago, IL 60637, USA}
\altaffiltext{5}{School of Physics, University of Exeter, EX4 4QL Exeter, UK}
\altaffiltext{6}{Department of Physics and Astronomy, University College London, Gower Street, WC1 E6BT, UK}
\altaffiltext{7}{Division of Geological and Planetary Sciences, California Institute of Technology, Pasadena, CA 91125, USA}

\begin{abstract}
We examine the repeatability, reliability, and accuracy of differential exoplanet eclipse depth measurements made using the InfraRed Array Camera (IRAC) on the {\sl Spitzer} Space Telescope during the post-cryogenic mission.  We have re-analyzed an existing 4.5\mum\ dataset, consisting of 10 observations of the XO-3b system during secondary eclipse, using seven different techniques for removing correlated noise.  We find that, on average, for a given technique the eclipse depth estimate is repeatable from epoch to epoch to within 156 parts per million (ppm).  Most techniques derive eclipse depths that do not vary by more than a factor 3 of the photon noise limit.  All methods but one accurately assess their own errors: for these methods the individual measurement uncertainties are comparable to the scatter in eclipse depths over the 10-epoch sample.  To assess the accuracy of the techniques as well as clarify the difference between instrumental and other sources of measurement error, we have also analyzed a simulated dataset of 10 visits to XO-3b, for which the eclipse depth is known.  We find that three of the methods (BLISS mapping, Pixel Level Decorrelation, and Independent Component Analysis) obtain results that are within 3 times the photon limit of the true eclipse depth.  When averaged over the 10-epoch ensemble, five out of seven techniques come within 100\,ppm of the true value.  \Spitzer\ exoplanet data, if measured following current best practices and reduced using methods such as those described here, can measure repeatable and accurate single eclipse depths, with close to photon-limited results.
\end{abstract}





\section{Introduction}
\subsection{Exoplanet measurements and correlated noise}
Measurement of relative flux variations is one of the chief means of characterizing transiting exoplanetary systems.  At infrared wavelengths secondary eclipses are a powerful tool for studying the atmospheres of giant exoplanets, their depths approximately equaling the dayside planet-to-star flux ratio.
Extracting information about atmospheres, however, is extremely challenging due to the small differential signals produced by transits, secondary eclipses, and phase curves.   The relevant signals are often at the level of 100 parts per million (ppm) or smaller, and require the removal of significant instrumental systematics in the two infrared instruments currently capable of providing information at this precision: the Wide Field Camera 3 (WFC3) on the {\sl Hubble Space Telecope} (HST) and the InfraRed Array Camera (IRAC, \citealt{2004ApJS..154...10F}) aboard the {\sl Spitzer} Space Telescope \citep{2004ApJS..154....1W}.  For the IRAC 3.6 and 4.5\,$\mu$m InSb detectors that remain active on post-cryogenic {\sl Spitzer}, the systematics are due to the interplay of residual telescope pointing fluctuations with intra-pixel gain variations in the moderately under-sampled camera.  

Over the past decade, a suite of techniques for removing time-correlated noise in IRAC data has been developed.  Because of the known coupling between pointing variations and the intra-pixel gain, the earliest methods for correcting cryogenic data used either a simple radial function from a pixel's center \citep{Reach:2005ji} or fit a second order polynomial to the observed flux variations as a function of the source centroid position \citep[e.g.,][]{2008ApJ...686.1341C}.  It soon became clear, however, that a single polynomial surface does not sufficiently describe the intra-pixel gain variations. To measure flux decrements with precision less than $\sim 1$\%, a more responsive approach is necessary to track the small-scale structure in the gain \citep{2010PASP..122.1341B}.  Furthermore, after {\sl Spitzer} entered its post-cryogenic stage in mid-2009, the amplitude of the variations doubled at the current detector temperature of about $28.7\,$K.\footnote{\url{http://irsa.ipac.caltech.edu/data/SPITZER/docs/irac/calibrationfiles/pixelphase/}}  

Thus, more flexible nonparametric approaches were developed to measure and remove the systematics.  The earliest such methods used some form of nearest neighbor kernel regression to map the intra-pixel gain as a function of centroid position, using a weighted sum of the measured fluxes instead of a predetermined function of centroid \citep{2010PASP..122.1341B}.  A special case of nearest neighbor kernel regression is BiLinearly Interpolated Subpixel Sensitivity (BLISS) mapping \citep{2012ApJ...754..136S}.  Additional promising techniques that have appeared in recent years include regression via Gaussian Processes \citep[GP;][]{2012MNRAS.419.2683G,2015MNRAS.451..680E}; Independent Component Analysis \citep[ICA;][]{2015ApJ...808...56M}; and Pixel Level Decorrelation \citep[PLD;][]{2015ApJ...805..132D}.  See Appendix B for a detailed review of these techniques. 

\subsection{Repeatability of Spitzer/IRAC Relative Flux Measurements}
As multi-epoch monitoring data have accumulated, investigators have begun to quantify the repeatability and reliability of exoplanet differential flux measurements made with \Spitzer\ and other observatories.  A growing body of evidence is showing that modern IRAC correlated noise removal techniques obtain consistent results from one measurement to the next, and obtain consistent results between techniques.  

One indicator of stability is that the individual measurement uncertainties approximately equal the scatter (standard deviation) in independently-measured transit or eclipse depths. For example, \citet{2013ApJ...765..127F} analyzed 14 transits of GJ 1214b measured at 4.5\mum\ with IRAC using a kernel regression decorrelation technique (KR/Data---see section B.4), yielding a scatter in transit depths within 50\% of the average reported uncertainty in the individual depths.  \citet{2014ApJ...794..134W} also used KR/Data to process data for 12 eclipses of XO-3b, yielding individual uncertainties that were equal to the scatter in the ensemble.  The XO-3b dataset features prominently in this paper, as a main component of the \Spitzer\ 2015 Data Challenge (see below).  

Older data often benefit from reanalysis with modern methods.  Four GJ 436b transits were reprocessed using ICA by \citet{2015ApJ...802..117M}, who determined that the transit depth did not vary by more than 100 ppm, contrary to earlier estimates computed using polynomial fitting \citep{2011ApJ...731...16B,2011ApJ...735...27K}.  The ICA technique was also used to establish a repeatable (within 200 ppm) transit depth for HD 189733b \citep{2014ApJ...786...22M}, after many conflicting prior values led to questions of stellar variability.  BLISS mapping \citep{2014ApJ...796...66D} and GP \citep{2015MNRAS.451..680E} were both used to reanalyze four eclipses of HD 209458b, including one taken under non-optimal observing conditions (see \S4.3).  Both teams concluded that the group of measurements was self-consistent  \citep[scatter 30\% less than uncertainties for][]{2015MNRAS.451..680E}, and that the earlier estimate of a much deeper occultation, which resulted in claims of a possible temperature inversion layer in the planet's atmosphere \citep{2008ApJ...673..526K}, was unwarranted.  

\subsection{Goals of this paper}
Because of the high relative precision required for eclipse depth and other exoplanet measurements, it is important to characterize the ability of an instrument---together with the chosen method of systematics removal---to return consistent results.  This is especially crucial when comparing data to models \citep[see][for a discussion of the difficulty of spectral retrieval from data with low S/N]{Burrows:2014jd} or measuring atmospheric variability \citep[e.g., see][who found evidence for eclipse depth changes of $\sim 140\,$ppm over 1 year in 55 Cnc e]{2016MNRAS.455.2018D}.  Despite the growing number of analyses of multi-epoch transit or eclipse measurements, all have thus far focused on at most two methods of removing correlated noise \citep{2013ApJ...765..127F,2014ApJ...794..134W,2015ApJ...802..117M,2014ApJ...796...66D,2015MNRAS.451..680E,2016MNRAS.455.2018D}, or only considered two epochs per target \citep{2014MNRAS.444.3632H}.  

This paper examines the repeatability of \Spitzer/IRAC eclipse depths in the post-cryogenic mission, with an eye towards answering the questions: how stable can we reasonably expect IRAC eclipse depth measurements to be; and how close are they to the truth?  We aim to establish limits on both the IRAC instrument and the best modern techniques for removing correlated noise and measuring eclipse depths, using both real and simulated data. Recently, participants undertook a Data Challenge consisting of the measurement of ten secondary eclipses of XO-3b \citep{2014ApJ...794..134W}, and a complementary analysis of a synthetic version of the XO-3b data.  In \S2 we describe the Data Challenge.  We introduce the real XO-3b dataset, give an overview of the {\sl Spitzer}/IRAC simulator and the creation of the simulated dataset, and outline seven techniques used to decorrelate the photometry.  In \S3 we report on the results of the data challenge, estimating the single eclipse depth repeatability and the reliability or precision of the results when reduced by the different methods. We compare the variability between methods, as well as the accuracy of the techniques when applied to simulated data.  In \S4 we discuss the implications of our results for post-cryogenic exoplanet measurements with \Spitzer.  We also evaluate a recent proposal to inflate IRAC eclipse depth uncertainties \citep{2014MNRAS.444.3632H}, and suggest application of our approach to future space observatories.  We conclude in \S5 by summarizing our key results.

\section{Methodology}
\subsection{The IRAC 2015 Data Challenge}
To assess the repeatability, reliability, and accuracy of post-cryogenic observations with IRAC, the {\sl Spitzer} Science Center (SSC) in conjunction with active exoplanet researchers from the astronomical community has performed an analysis of the removal of systematics and measured the repeatability of warm IRAC observations.  The SSC made available to the public both a real dataset as well as synthetic data (where the eclipse depth is an input) on the IRAC Data Challenge 2015 website\footnote{\url{http://irachpp.spitzer.caltech.edu/page/data-challenge-2015}}.  Contributions were solicited, and preliminary results were presented at the IRAC 2nd Workshop on High Precision Photometry, held during the 2015 International Astronomical Union meeting in Honolulu, HI, USA\footnote{\url{http://irachpp.spitzer.caltech.edu/page/IRAC\_IAU\_2015}}.  In this section we describe the real and simulated data and the decorrelation techniques used. 

\begin{deluxetable*}{cccccccrr}
\tablecaption{Real {\sl Spitzer} XO-3b Eclipse AORs and Positions\label{xo3meas}}
\tablewidth{0pt}
\tablehead{
\colhead{Start Time\tablenotemark{a}} & \multicolumn{2}{c}{AOR Number\tablenotemark{b}} & \colhead{$\langle X\rangle$\tablenotemark{c}} & \colhead{$\sigma_X$\tablenotemark{d}} & \colhead{$\langle Y\rangle$\tablenotemark{c}} &
\colhead{$\sigma_Y$\tablenotemark{d}} & \colhead{$\sigma_{XY}$\tablenotemark{e}} & \colhead{No.\tablenotemark{f}} \\
\colhead{(JD-2455000)} &Pre&Main&
\colhead{(px)} & \colhead{(px)} &
\colhead{(px)} & \colhead{(px)} &\colhead{($10^{-4}\,$px$^2$)} & \colhead{}
}\colnumbers
\startdata
1242.2402 & \dataset[46467072]{http://sha.ipac.caltech.edu/applications/Spitzer/SHA/\#id=SearchByRequestID\&RequestClass=ServerRequest\&DoSearch=true&SearchByRequestID.field.requestID=46467072\&SearchByRequestID.field.includeSameConstraints=\_none\_\&MoreOptions.field.prodtype=aor\&shortDesc=AORKEY\&isBookmarkAble=true\&isDrillDownRoot=true\&isSearchResult=true}
& \dataset[46471424]{http://sha.ipac.caltech.edu/applications/Spitzer/SHA/\#id=SearchByRequestID\&RequestClass=ServerRequest\&DoSearch=true&SearchByRequestID.field.requestID=46471424\&SearchByRequestID.field.includeSameConstraints=\_none\_\&MoreOptions.field.prodtype=aor\&shortDesc=AORKEY\&isBookmarkAble=true\&isDrillDownRoot=true\&isSearchResult=true}
& 15.17 & 0.03 & 15.00 & 0.05 & --8.56\phantom{000} &  2 \\
1248.6482 & \dataset[46467840]{http://sha.ipac.caltech.edu/applications/Spitzer/SHA/\#id=SearchByRequestID\&RequestClass=ServerRequest\&DoSearch=true&SearchByRequestID.field.requestID=46467840\&SearchByRequestID.field.includeSameConstraints=\_none\_\&MoreOptions.field.prodtype=aor\&shortDesc=AORKEY\&isBookmarkAble=true\&isDrillDownRoot=true\&isSearchResult=true} & \dataset[46471168]{http://sha.ipac.caltech.edu/applications/Spitzer/SHA/\#id=SearchByRequestID\&RequestClass=ServerRequest\&DoSearch=true&SearchByRequestID.field.requestID=46471168\&SearchByRequestID.field.includeSameConstraints=\_none\_\&MoreOptions.field.prodtype=aor\&shortDesc=AORKEY\&isBookmarkAble=true\&isDrillDownRoot=true\&isSearchResult=true} & 15.10 & 0.04 & 15.14 & 0.06 & 9.53\phantom{000} &  3 \\
1251.8187 & \dataset[46470144]{http://sha.ipac.caltech.edu/applications/Spitzer/SHA/\#id=SearchByRequestID\&RequestClass=ServerRequest\&DoSearch=true&SearchByRequestID.field.requestID=46470144\&SearchByRequestID.field.includeSameConstraints=\_none\_\&MoreOptions.field.prodtype=aor\&shortDesc=AORKEY\&isBookmarkAble=true\&isDrillDownRoot=true\&isSearchResult=true} & \dataset[46470912]{http://sha.ipac.caltech.edu/applications/Spitzer/SHA/\#id=SearchByRequestID\&RequestClass=ServerRequest\&DoSearch=true&SearchByRequestID.field.requestID=46470912\&SearchByRequestID.field.includeSameConstraints=\_none\_\&MoreOptions.field.prodtype=aor\&shortDesc=AORKEY\&isBookmarkAble=true\&isDrillDownRoot=true\&isSearchResult=true} & 15.23 & 0.06 & 15.03 & 0.06 & --25.63\phantom{000} &  4 \\
1255.0166 & \dataset[46467584]{http://sha.ipac.caltech.edu/applications/Spitzer/SHA/\#id=SearchByRequestID\&RequestClass=ServerRequest\&DoSearch=true&SearchByRequestID.field.requestID=46467584\&SearchByRequestID.field.includeSameConstraints=\_none\_\&MoreOptions.field.prodtype=aor\&shortDesc=AORKEY\&isBookmarkAble=true\&isDrillDownRoot=true\&isSearchResult=true} & \dataset[46470656]{http://sha.ipac.caltech.edu/applications/Spitzer/SHA/\#id=SearchByRequestID\&RequestClass=ServerRequest\&DoSearch=true&SearchByRequestID.field.requestID=46470656\&SearchByRequestID.field.includeSameConstraints=\_none\_\&MoreOptions.field.prodtype=aor\&shortDesc=AORKEY\&isBookmarkAble=true\&isDrillDownRoot=true\&isSearchResult=true} & 15.17 & 0.04 & 15.13 & 0.05 & 4.25\phantom{000} &  5 \\
1264.5897 & \dataset[46469376]{http://sha.ipac.caltech.edu/applications/Spitzer/SHA/\#id=SearchByRequestID\&RequestClass=ServerRequest\&DoSearch=true&SearchByRequestID.field.requestID=46469376\&SearchByRequestID.field.includeSameConstraints=\_none\_\&MoreOptions.field.prodtype=aor\&shortDesc=AORKEY\&isBookmarkAble=true\&isDrillDownRoot=true\&isSearchResult=true} & \dataset[46470400]{http://sha.ipac.caltech.edu/applications/Spitzer/SHA/\#id=SearchByRequestID\&RequestClass=ServerRequest\&DoSearch=true&SearchByRequestID.field.requestID=46470400\&SearchByRequestID.field.includeSameConstraints=\_none\_\&MoreOptions.field.prodtype=aor\&shortDesc=AORKEY\&isBookmarkAble=true\&isDrillDownRoot=true\&isSearchResult=true} & 15.19 & 0.03 & 14.99 & 0.06 & --4.31\phantom{000} &  6 \\
1270.9776 & \dataset[46466816]{http://sha.ipac.caltech.edu/applications/Spitzer/SHA/\#id=SearchByRequestID\&RequestClass=ServerRequest\&DoSearch=true&SearchByRequestID.field.requestID=46466816\&SearchByRequestID.field.includeSameConstraints=\_none\_\&MoreOptions.field.prodtype=aor\&shortDesc=AORKEY\&isBookmarkAble=true\&isDrillDownRoot=true\&isSearchResult=true} & \dataset[46469632]{http://sha.ipac.caltech.edu/applications/Spitzer/SHA/\#id=SearchByRequestID\&RequestClass=ServerRequest\&DoSearch=true&SearchByRequestID.field.requestID=46469632\&SearchByRequestID.field.includeSameConstraints=\_none\_\&MoreOptions.field.prodtype=aor\&shortDesc=AORKEY\&isBookmarkAble=true\&isDrillDownRoot=true\&isSearchResult=true} & 15.13 & 0.06 & 15.12 & 0.05 & 9.22\phantom{000} &  7 \\ 
1405.0165 & \dataset[46468864]{http://sha.ipac.caltech.edu/applications/Spitzer/SHA/\#id=SearchByRequestID\&RequestClass=ServerRequest\&DoSearch=true&SearchByRequestID.field.requestID=46468864\&SearchByRequestID.field.includeSameConstraints=\_none\_\&MoreOptions.field.prodtype=aor\&shortDesc=AORKEY\&isBookmarkAble=true\&isDrillDownRoot=true\&isSearchResult=true} & \dataset[46469120]{http://sha.ipac.caltech.edu/applications/Spitzer/SHA/\#id=SearchByRequestID\&RequestClass=ServerRequest\&DoSearch=true&SearchByRequestID.field.requestID=46469120\&SearchByRequestID.field.includeSameConstraints=\_none\_\&MoreOptions.field.prodtype=aor\&shortDesc=AORKEY\&isBookmarkAble=true\&isDrillDownRoot=true\&isSearchResult=true} & 15.21 & 0.03 & 14.92 & 0.04 & --3.93\phantom{000} &  8 \\
1430.5523 & \dataset[46469888]{http://sha.ipac.caltech.edu/applications/Spitzer/SHA/\#id=SearchByRequestID\&RequestClass=ServerRequest\&DoSearch=true&SearchByRequestID.field.requestID=46469888\&SearchByRequestID.field.includeSameConstraints=\_none\_\&MoreOptions.field.prodtype=aor\&shortDesc=AORKEY\&isBookmarkAble=true\&isDrillDownRoot=true\&isSearchResult=true} & \dataset[46468608]{http://sha.ipac.caltech.edu/applications/Spitzer/SHA/\#id=SearchByRequestID\&RequestClass=ServerRequest\&DoSearch=true&SearchByRequestID.field.requestID=46468608\&SearchByRequestID.field.includeSameConstraints=\_none\_\&MoreOptions.field.prodtype=aor\&shortDesc=AORKEY\&isBookmarkAble=true\&isDrillDownRoot=true\&isSearchResult=true} & 15.15 & 0.03 & 15.01 & 0.05 & --4.45\phantom{000} &  10 \\  
1433.7433 & \dataset[46467328]{http://sha.ipac.caltech.edu/applications/Spitzer/SHA/\#id=SearchByRequestID\&RequestClass=ServerRequest\&DoSearch=true&SearchByRequestID.field.requestID=46467328\&SearchByRequestID.field.includeSameConstraints=\_none\_\&MoreOptions.field.prodtype=aor\&shortDesc=AORKEY\&isBookmarkAble=true\&isDrillDownRoot=true\&isSearchResult=true} & \dataset[46468352]{http://sha.ipac.caltech.edu/applications/Spitzer/SHA/\#id=SearchByRequestID\&RequestClass=ServerRequest\&DoSearch=true&SearchByRequestID.field.requestID=46468352\&SearchByRequestID.field.includeSameConstraints=\_none\_\&MoreOptions.field.prodtype=aor\&shortDesc=AORKEY\&isBookmarkAble=true\&isDrillDownRoot=true\&isSearchResult=true} & 15.21 & 0.03 & 14.99 & 0.05 & --4.04\phantom{000} &  11 \\
1436.9273 & \dataset[46471680]{http://sha.ipac.caltech.edu/applications/Spitzer/SHA/\#id=SearchByRequestID\&RequestClass=ServerRequest\&DoSearch=true&SearchByRequestID.field.requestID=46471680\&SearchByRequestID.field.includeSameConstraints=\_none\_\&MoreOptions.field.prodtype=aor\&shortDesc=AORKEY\&isBookmarkAble=true\&isDrillDownRoot=true\&isSearchResult=true} & \dataset[46468096]{http://sha.ipac.caltech.edu/applications/Spitzer/SHA/\#id=SearchByRequestID\&RequestClass=ServerRequest\&DoSearch=true&SearchByRequestID.field.requestID=46468096\&SearchByRequestID.field.includeSameConstraints=\_none\_\&MoreOptions.field.prodtype=aor\&shortDesc=AORKEY\&isBookmarkAble=true\&isDrillDownRoot=true\&isSearchResult=true} & 15.23 & 0.04 & 14.96 & 0.05 & --2.80\phantom{000} &  12 \\
\hline
\multicolumn{3}{r}{Column Means\tablenotemark{g}} & 15.18 & 0.04 & 15.03 & 0.05 & --3.07\phantom{000} & \\
\multicolumn{3}{r}{All Data\tablenotemark{h}} & 15.18 & 0.06 & 15.03 & 0.09 & --26.63\phantom{000} & \\
\enddata
\tablenotetext{a}{Start time of first exposure of initial AOR.}
\tablenotetext{b}{Electronic versions of this table contain links to these datasets in the {\sl Spitzer} archive.}
\tablenotetext{c}{Mean centroid over all measurements in the two AORs.}
\tablenotetext{d}{Standard deviation in centroid over all measurements in the two AORs.}
\tablenotetext{e}{$(x,y)$ Covariance in centroid over all measurements in the two AORs.}
\tablenotetext{f}{Eclipse number as listed in Table 1 of \citet{2014ApJ...794..134W} (not all eclipses analyzed by \citeauthor{2014ApJ...794..134W} were part of the Data Challenge).}
\tablenotetext{g}{Mean, standard deviation, and $(x,y)$ covariance of centroid averaged along the table column.}
\tablenotetext{h}{Mean, standard deviation, and $(x,y)$ covariance of centroid over all AORs.}
\end{deluxetable*}

\subsection{Real XO-3b Observations}
The XO-3b data used for the Data Challenge consisted of 10 individual secondary eclipse measurements originally analyzed by \citet{2014ApJ...794..134W}, and summarized in Table 1.  All measurements were made with post-cryogenic \Spitzer\ in 2012 and 2013, and were taken as part of Program ID (PID) 90032 (PI: H. Knutson).  This program also contains two full phase curve measurements of XO-3b at 4.5\mum, but we confine our analysis in this paper to the eclipse--only datasets. The first six epochs took place within about 30 days of each other; the last four occurred about one-half year later and also spanned 30 days.  Each epoch consisted of two Astronomical Observation Requests (AORs): an 11-exposure, 30 minute ``Pre'' AOR to allow short term pointing drift to settle; and a 233-exposure, 8.5 hour ``Main'' AOR that contained the secondary eclipse.  Each exposure produced a FITS format image file, containing a cube of 64 32$\times$32 pixel images taken 2\,seconds apart with the source in the subarray field-of-view on the 4.5\,$\mu$m array.  The measurements were taken in staring mode (no repositioning within an AOR), and used PCRS Peak-Up to establish the position of XO-3b at the beginning of each AOR.\footnote{\url{http://irachpp.spitzer.caltech.edu/page/Obs\%20Planning}} Table \ref{xo3meas} gives the observation start time, {\sl Spitzer} AOR numbers, and the eclipse number \citep[for comparison with][Table 1, which also includes two full phase curve datasets]{2014ApJ...794..134W}.

\begin{deluxetable*}{cccccccr}
\tablecaption{Synthetic {\sl Spitzer} XO-3b AORs and Positions\label{sim_xo3meas}}
\tablewidth{0pt}
\tablehead{
\colhead{Start Time\tablenotemark{a}} & \multicolumn{2}{c}{AOR Number\tablenotemark{b}} & \colhead{$\langle X\rangle$} & \colhead{$\sigma_X$} & \colhead{$\langle Y\rangle$} &
\colhead{$\sigma_Y$} & \colhead{$\sigma_{XY}$}\\
\colhead{(JD-2455000)} &Pre&Main&
\colhead{(px)} & \colhead{(px)} &
\colhead{(px)} & \colhead{(px)} &\colhead{($10^{-4}\,$px$^2$)} 
}\colnumbers
\startdata
2206.1459 & 20150000 & 20150001 & 15.16 & 0.02 & 15.01 & 0.05 & --5.93\phantom{000} \\
2209.2927 & 20150002 & 20150003 & 15.12 & 0.02 & 15.08 & 0.03 & --4.57\phantom{000} \\
2212.5079 & 20150004 & 20150005 & 15.16 & 0.03 & 15.13 & 0.05 & --10.33\phantom{000} \\
2215.7214 & 20150006 & 20150007 & 15.10 & 0.02 & 15.06 & 0.10 & --16.34\phantom{000} \\
2218.9047 & 20150008 & 20150009 & 15.20 & 0.03 & 15.17 & 0.07 & --10.95\phantom{000} \\
2222.0547 & 20150010 & 20150011 & 15.12 & 0.02 & 15.17 & 0.04 & --5.19\phantom{000} \\
2225.2356 & 20150012 & 20150013 & 15.18 & 0.03 & 15.05 & 0.09 & --20.87\phantom{000} \\
2228.4898 & 20150014 & 20150015 & 15.17 & 0.02 & 15.09 & 0.10 & --10.61\phantom{000} \\
2231.6296 & 20150016 & 20150017 & 15.14 & 0.02 & 15.17 & 0.05 & --7.41\phantom{000} \\
2234.8406 & 20150018 & 20150019 & 15.10 & 0.03 & 15.09 & 0.06 & --6.71\phantom{000} \\
\hline
\multicolumn{3}{r}{Column Means} & 15.15 & 0.03 & 15.10 & 0.06 & --9.90\phantom{000} \\
\multicolumn{3}{r}{All Data} & 15.15 & 0.04 & 15.10 & 0.09 & --8.64\phantom{000}\\ 
\enddata
\tablenotetext{a}{Simulated start time of first exposure of initial AOR.}
\tablenotetext{b}{Data may be downloaded from \url{http://irachpp.spitzer.caltech.edu/page/data-challenge-2015}}
\end{deluxetable*}

\subsection{Synthetic XO-3b Observations}
Observed variations in eclipse depths are caused by a combination of variations in \Spitzer\ pointing, IRAC detector charge trapping, and possible evolution of the planetary system, as well as the limitations and biases of the technique for reducing correlated noise.  We can analyze the data using different techniques to assess differences in the methods, but it is often difficult with real data to completely separate pointing from instrumental or planetary variations.  This is one reason we have included synthetic data as part of the Data Challenge, for which both the exoplanet and IRAC are given constant properties.  We had originally considered using eclipsing binary stars observed with Kepler as a truth set which could then be observed with \Spitzer.  Unfortunately, using stellar atmosphere models to extrapolate Kepler eclipse depths to \Spitzer\ wavelengths are as fraught with potential uncertainty as the planetary eclipse depths themselves, suggesting simulated data are the only reasonable path to estimating accuracy.  In the simulations, any measured variations in eclipse depth are due solely to (1) random noise and (2) residual correlated noise not removed by decorrelation analysis.  This should give us a better insight into the capabilities of the decorrelation methods than real data alone. 

To produce the simulated XO-3b observations used for the Data Challenge, we used {\tt IRACSIM}, a package built in the IDL programming language.  The program uses a model of the \Spitzer/IRAC system to create synthetic IRAC point source measurements, outputting FITS image (or image cube) files similar to those produced by the IRAC basic calibrated data (BCD) pipeline. We give an overview of the model in Appendix A.

Table \ref{sim_xo3meas} gives the simulated observation start times and AOR numbers of the synthetic observations.  The simulations followed closely the design of the real observations, with each observing ``epoch'' containing 2 AORs, a similar number of exposures per AOR, and the same integration parameters.  We set the start times for each synthetic epoch at slightly different phases of different actual XO-3b orbits, as often occurs in real observations.  This allows for different proportions of samples before and after eclipse for each epoch to minimize biases in fitting.

Table 
\ref{ptg_params} 
lists the range of inputs to the pointing model used in simulating the XO-3b data. We chose not to duplicate exactly the pointing fluctuations as observed in the real dataset, but attempted to simulate a range of possible \Spitzer\ observing conditions (drawn roughly from the distribution of observed cases), and thus a range of possible decorrelation situations.  In practice this resulted in generally larger pointing fluctuations and drifts than found in the real data.

\begin{figure}
\plotone{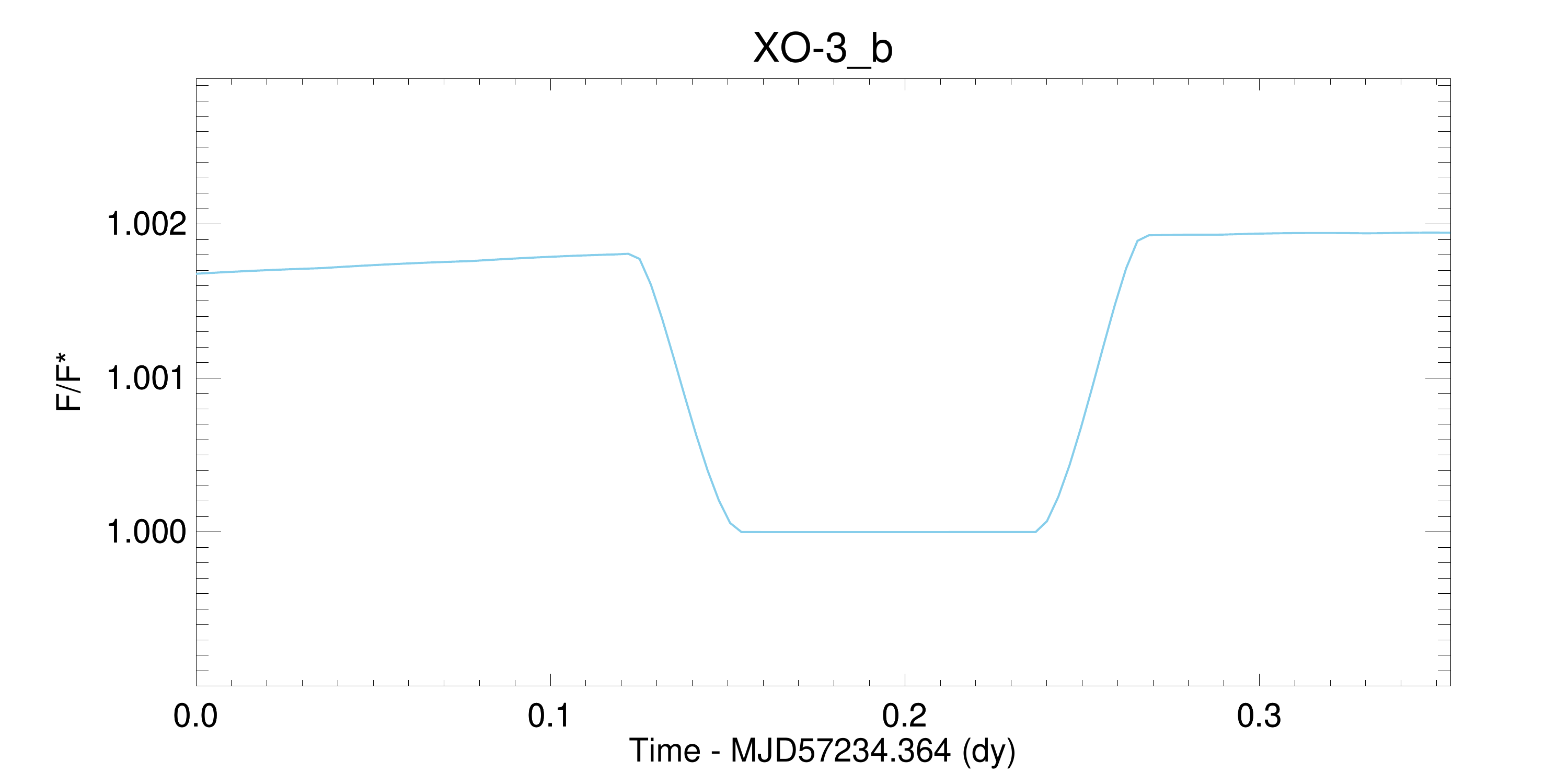}
\caption{Light curve for simulated XO-3b observations, AOR number 20150019.  The simulation uses known values of the system's stellar, orbital, and transit parameters, in addition to a thermal phase model with albedo, $A=0$; radiative timescale, $\tau_{\litl rad}=1\,$day; and net rotational angular velocity of the cloud layer, $\Omega_{\litl rot} = 1$, in units of the orbital angular velocity at periastron.  The eclipse depth is 1875\,ppm.
\label{predicted_lightcurve}
}
\end{figure}

We used the {\tt IRACSIM} exoplanet wrapper to model the light curve of XO-3b, obtaining values for the system's stellar, orbital, and transit parameters from the exoplanets.org database (as of 2015 July 2) and simulating the planet's thermal phase variations using the model of \citet{2011ApJ...726...82C}.  Since the goal was to understand IRAC data, not XO-3b, we set somewhat arbitrary values of the planetary parameters: (1) albedo, $A=0$; (2) radiative timescale, $\tau_{\litl rad}=1\,$day; and (3) net rotational angular velocity of the cloud layer, $\Omega_{\litl rot} = 1$ (in units of the orbital angular velocity at periastron).  The resulting phase curve gives a non-flat appearance to the flux outside of eclipse and sets the depth of the eclipse, which we define in terms of the stellar flux.  In this case, the model eclipse depth for XO-3b is 1875\,ppm, about 16\% larger than the actual depth published by \citet{2014ApJ...794..134W}.  The model light curve for the 10th epoch is shown in Figure \ref{predicted_lightcurve}.

\subsection{Decorrelation techniques}
The best hypothesis for the source of IRAC time-correlated noise is the coupling of pointing fluctuations with intrapixel quantum efficiency variations on the InSb detector arrays.  When \Spitzer\ is commanded to continuously observe an inertially fixed target (``staring'' mode), a source position will undergo ``jitter'' and ``wobble'' with a net amplitude of about 0.08 detector pixels (px) per hour, while also incurring a slow linear drift of about 0.01\,px per hour (see \S A.1 for an analytical model of these fluctuations). These telescope motions have been described in detail by \citet{Grillmair:2012fq}, and the physical causes of some are known.  For example, the wobble is caused by a battery heater cycling on and off with period of $\sim 40\,$min, and the long term drift ($y$ pixel direction only) is caused by the discrepancy between the instantaneous velocity aberration of the spacecraft and the on-board aberration correction that occurs only at the start of an AOR.  A map of the photometric gain of a point source on the central pixel of the 4.5\mum\ subarray is displayed in Figure \ref{xo3b_pos_real}, showing that correlated noise due to pointing fluctuations can be as much as 1--2\%, a factor of 10 larger than the XO-3b eclipse depth.

As part of the Data Challenge, exoplanet experts used a total of seven different data reduction techniques to remove correlated noise from the \Spitzer/IRAC photometry and assess the eclipse depth repeatability.  We review the seven techniques in Appendix B, including notes on implementation for the XO-3b datasets.  Among these are the most commonly used techniques in the current literature to date (BLISS, KR/Data), as well as a group of more recently developed methods [GP, ICA, KR/Pmap, PLD, Segmented Polynomial (K2 pipeline)].  Note that each expert was free to use any approach to centroiding, photometry, and eclipse depth fitting.  Thus, any mention of a method by name in this paper refers to the {\it entire data reduction pipeline}, not just the correlated noise removal algorithm. 

\section{Results}
\subsection{XO-3b Centroids and Photometry}
\begin{figure}
\plotone{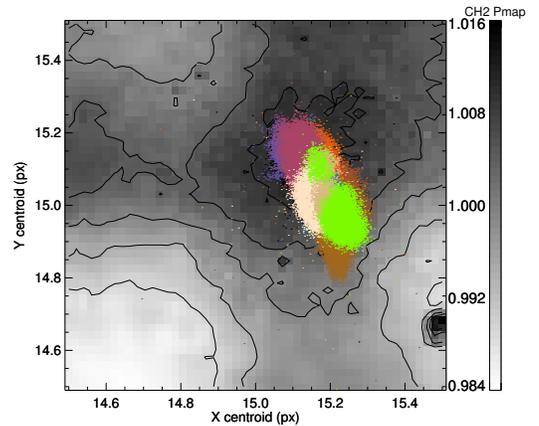}
\caption{Centroid positions (derived using the center-of-light method) of XO-3 on the IRAC 4.5\,$\mu$m subarray, from the {\bf real} dataset.  Each colored group of points indicates a separate epoch of observation (see Table \ref{xo3meas} for details on the epochs).  The background grayscale and contours shows the intra-pixel photometric gain map (``Pmap''), as measured using kernel regression on a calibration star (\S\ref{KRsection}).  The geometric center of the pixel is located at coordinates (15.0,15.0). \label{xo3b_pos_real}}
\end{figure}
\begin{figure}
\plotone{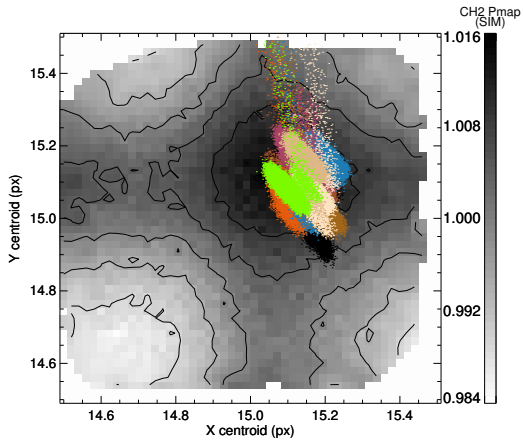}
\caption{Centroid positions (derived using the center-of-light method) of XO-3 on the IRAC 4.5\,$\mu$m subarray, from the {\bf simulated} dataset.  Each colored group of points indicates a separate epoch of observation (see Table \ref{sim_xo3meas} for details on the simulated epochs).  The background grayscale and contours shows the intra-pixel photometric gain map (``Pmap''), as measured using kernel regression on a simulated calibration star (\S\ref{KRsection}).\label{xo3b_pos_fake}}
\end{figure}

We begin with an overview of the data characteristics.  Figure \ref{xo3b_pos_real} plots all centroid positions for the individual measurements on the subarray center pixel for the real data, and Figure \ref{xo3b_pos_fake} does the same for the simulated data.  Because of the dependence of correlated noise on pointing fluctuations, most noise-removal techniques use source centroid as a primary decorrelation variable.  Most techniques described in \S2.4 use either 2D Gaussian fitting  or the flux-weighted ``center of light'' method to determine a point source center on the undersampled {\sl Spitzer} arrays. In this paper, if not stated otherwise, we only report center of light centroids:
\begin{eqnarray}
x_c & = & \frac{\sum_i i\,(f_{ij}-f_{\litl BG})}{\sum_i (f_{ij}-f_{\litl BG})}; \label{x_centroid}\\
y_c & = & \frac{\sum_j j\,(f_{ij}-f_{\litl BG})}{\sum_j (f_{ij}-f_{\litl BG})}\label{y_centroid}. 
\end{eqnarray}
Here, ($i,j$) is the pixel number, $f_{ij}$ is the image value at that pixel, and $f_{\litl BG}$ is the background flux in surrounding pixels.  The sums are over a ($7\times 7$)-pixel region surrounding the expected position of the source.   This centroiding method is sufficiently precise for decorrelation, resulting in positional distortions of at most 0.05 pixels \citep{2014SPIE.9143E..1MI}. A detailed discussion of different centroiding techniques is beyond the scope of this paper; see \citet{2014PASP..126.1092L} for analysis of the accuracy of three centroiding methods.

Columns 4--8 of Tables \ref{xo3meas} (real) and \ref{sim_xo3meas} (simulated) summarize the centroid ``clouds'' for each epoch, giving the means and standard deviations in $x$ and $y$ position, as well as the $xy$ covariances in centroid.  As the real data were all pointed using PCRS peak-up, the mean positions are all within 0.4 pixel of one another, and cluster near the peak of the intra-pixel gain.  Negative covariances for most of the real AORs indicate that the clouds are aligned such that $y$ decreases when $x$ increases, which is a common direction for \Spitzer/IRAC short-term drift.  The bottom 2 rows of Tables \ref{xo3meas} and \ref{sim_xo3meas} list the column means and the statistics for all data taken together.  For the real data, the full dataset has a much higher negative covariance than the individual clouds, suggesting that separate pointings fall preferentially along a $\sim -45\arcdeg$ axis.  The simulated data feature a much stronger initial drift, as well as a more pronounced $y$ component to the jitter, wobble, and drift than the real data.  Some individual $xy$ covariances in the simulated data are much higher than both the mean and the aggregate covariance for the group, due to this exaggerated elongation.  This ``stretching'' of the positions along $y$ reduces the positional redundancy and, as we will see, challenges the ability of most reduction methods to decorrelate the data.  

\begin{figure*}
\plotone{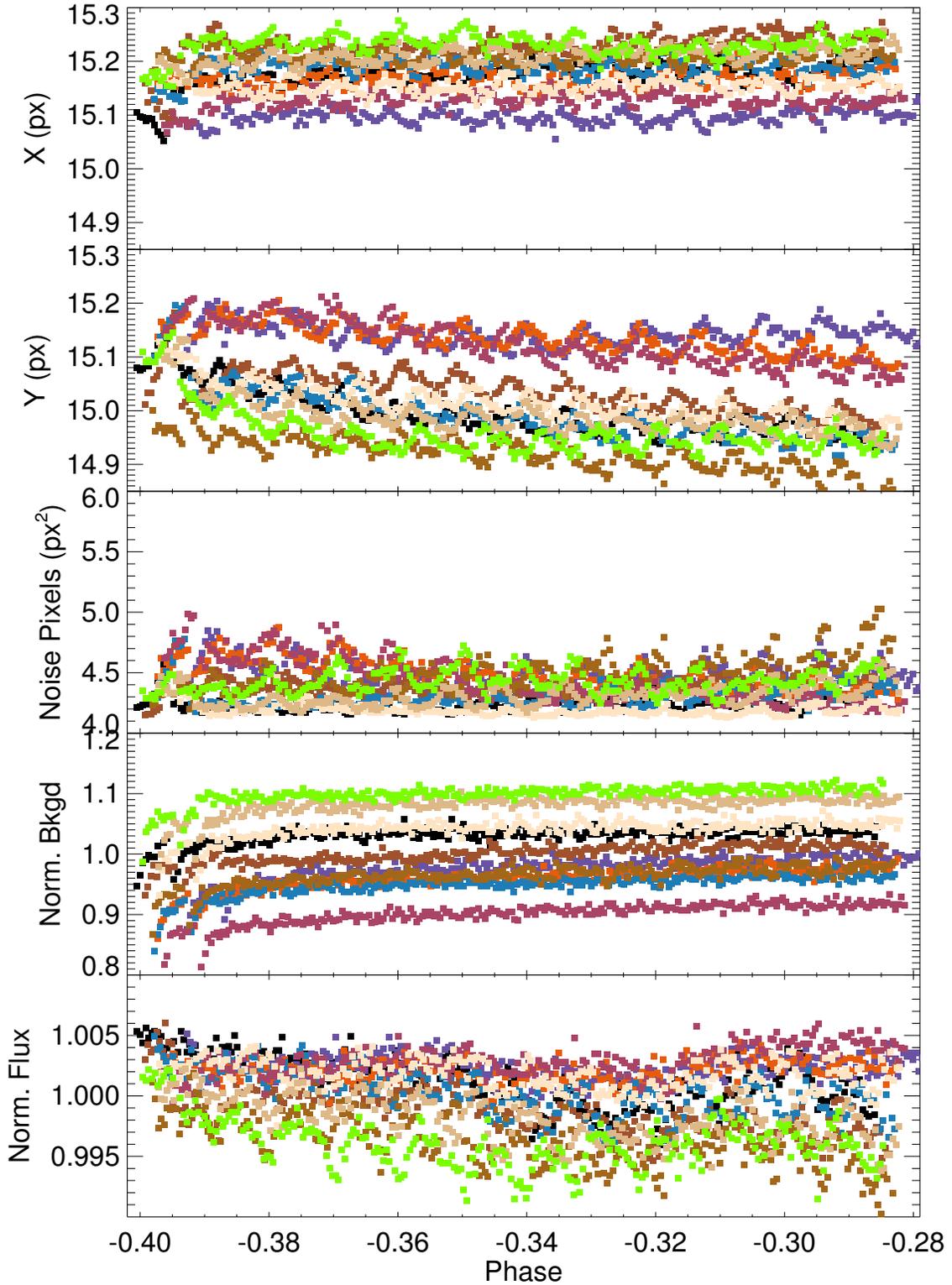}
\caption{{\bf Real} XO-3b photometric and other measurements as a function of orbital phase (fraction of orbit since transit).  From top to bottom: $x$ and $y$ centroid positions; Noise Pixel parameter, $\tilde{\beta}$; photometric background in 3--7 pixel radius annulus surrounding centroid, normalized to the mean over all AORs; and photometric flux in 2.25-pixel radius aperture, normalized to the mean over all AORs. Each point on the plots is the average of 63 measurements, or $\sim 2\,$min of integration.  We drop the first frame of each 64-frame subarray cube, to minimize residual bias pattern effects (the ``first frame effect''; see \url{http://irsa.ipac.caltech.edu/data/SPITZER/docs/irac/features/\#1}).  Each colored group of points indicates a separate epoch of observation (see Table \ref{xo3meas} for details on the epochs).  The time span is about 9 hours. \label{xo3b_phase_real}}
\end{figure*}
\begin{figure*}
\plotone{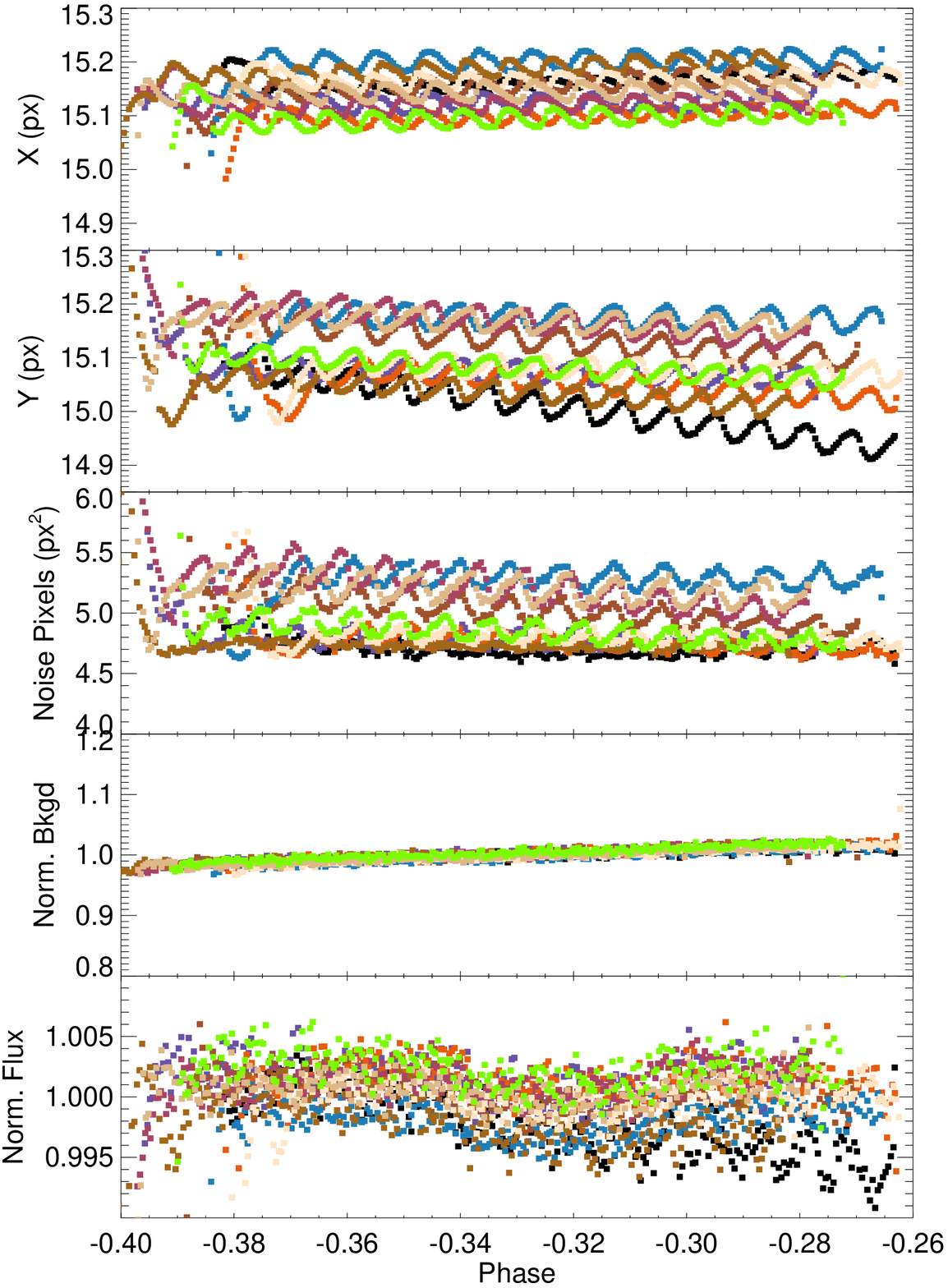}
\caption{{\bf Simulated} XO-3b photometric and other measurements as a function of orbital phase.  See caption to Fig. \ref{xo3b_phase_real} for description.   Each colored group of points indicates a separate epoch of observation (see Table \ref{sim_xo3meas} for details on the epochs).  Vertical scales for each panel are identical to those in Fig. \ref{xo3b_phase_real}. \label{xo3b_phase_fake}}
\end{figure*}

We display the real XO-3b photometric and other measurements as a function of orbital phase in Figure \ref{xo3b_phase_real}, and those for the synthetic data in Figure \ref{xo3b_phase_fake}.   As mentioned, some decorrelation methods use the Noise Pixels parameter, 
\begin{equation}
\tilde{\beta} = \frac{(\sum_{ij}f_{ij})^2}{\sum_{ij} f_{ij}^2},
\label{noise_pixels}
\end{equation}
which approximates the effective area (in square pixels) of a point source.  The sums are over the same $(7\times 7)$-pixel region over which the centroid is derived.  We display $\tilde{\beta}$ as a function of phase in the third panel of Figures \ref{xo3b_phase_real} and \ref{xo3b_phase_fake}.  This parameter partly measures observing geometry:  given constant total flux---the numerator of Equation \ref{noise_pixels} does not change---moving a source from the center to the edge of a pixel will spread the light to more pixels, decreasing the denominator and increasing $\tilde{\beta}$.  Thus we see in Figures \ref{xo3b_phase_real} and \ref{xo3b_phase_fake} how $\tilde{\beta}$ is correlated with the centroids to an extent. But $\tilde{\beta}$ also can measure the smearing of the IRAC PSF due to changes in the amplitude of high frequency jitter.  \Spitzer\ is known to have normal modes of oscillation with period less than the detector sample time (see \S{A.3} for a discussion of IRAC sampling).  When the amplitude of oscillation changes, the centroid might not vary markedly but the integrated PSF will change its apparent size, altering $\tilde{\beta}$.

We display the mean background in the fourth panel, and the aperture flux in the fifth panel of Figures \ref{xo3b_phase_real} and \ref{xo3b_phase_fake}.  We normalized the values in these last two panels to the mean value over all AORs, allowing us to notice relative shifts between AORs.  Fluxes are estimated using the IDL photometry program {\tt aper.pro}, with a 2.25-pixel radius circular aperture and a 3--7 pixel background annulus. Backgrounds are the mean value per pixel in the annulus, scaled to the area of the aperture. The net aperture flux is thus the integrated intensity per pixel weighted by the fraction of each pixel lying inside the aperture, minus the background value.  (Each team participating in the Data Challenge may have used a different method for measuring the  flux, including different aperture sizes or background definitions.)

\begin{figure*}
\plotone{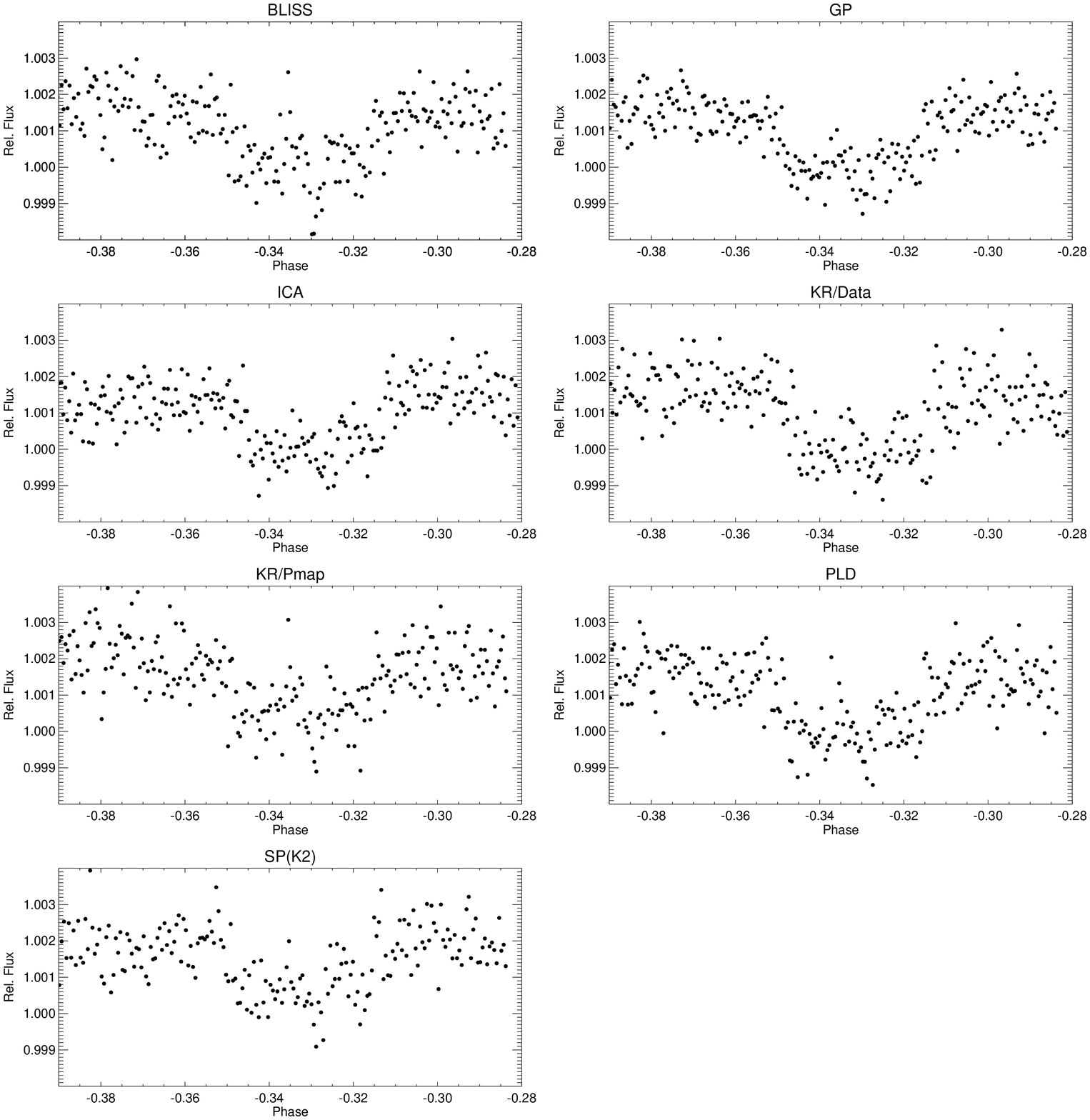}
\caption{Decorrelated light curves for {\bf real} XO-3b measurements, 5th epoch (\Spitzer\ AOR number 46470400).  Fluxes have been binned $64\times$.
\label{lightcurves_real}}
\end{figure*}

\begin{figure*}
\plotone{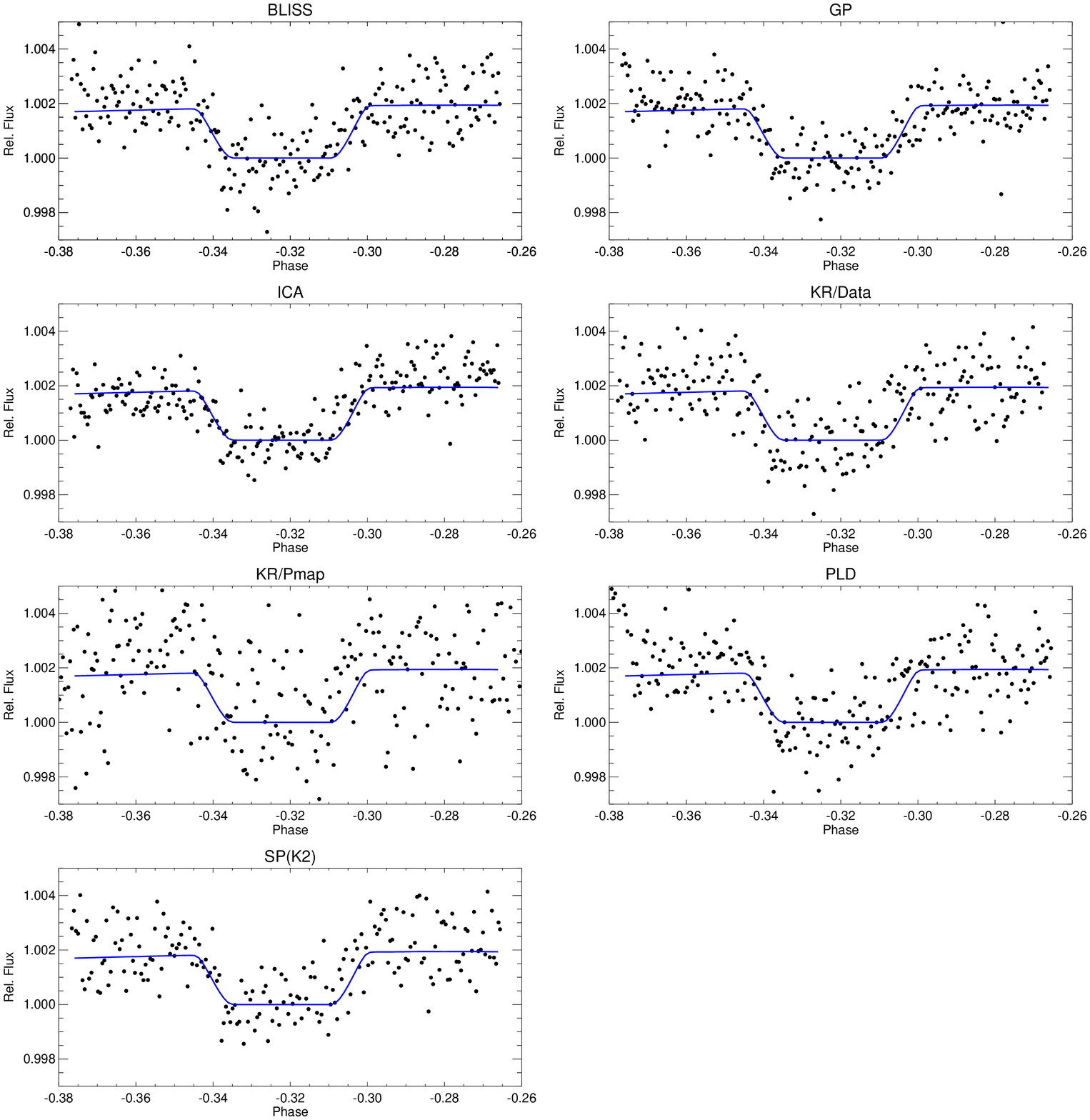}
\caption{Decorrelated light curves for {\bf simulated} XO-3b measurements, 5th epoch (simulated AOR number 20150009).  Fluxes have been binned $64\times$.  We overlay the input model lightcurve (not a fit) as a blue solid line.
\label{lightcurves_sim}}
\end{figure*}

Various known features of \Spitzer\ data can be seen in Figure \ref{xo3b_phase_real}.  The short term pointing drift, as well as the sawtooth-shaped ``wobble'', can be seen in the $x$ and $y$ centroids (and to a lesser extent in $\tilde{\beta}$; pointing effects are stronger in the $y$ direction).  The aperture fluxes for some epochs show very clearly the correlated noise signature due to these telescope motions. The eclipse can also be seen in the flux between phase --0.35 and --0.31.  

The background values show a quick ramp in the beginning of each epoch, and settle into a much slower increase with time for the final eight or so hours.  This behaves similarly to the ``flux ramp'' seen by many who work on 4.5\mum\ staring mode IRAC data \citep[e.g.,][]{2012ApJ...754...22K,2013ApJ...766...95L}.  In this case, however, the ramp disappears after background subtraction so the background ramp is probably caused by a relaxation in detector bias (the IRAC dark bias has a significant well-known offset that changes with time based on the history of readouts and array idling over the previous several hours), not changing responsivity.  The background curves in Fig. \ref{xo3b_phase_real} are all normalized to the same value; the fact that they are separated suggests a different mean background between epochs.  This can be attributed to fluctuations in the mean detector dark bias or changes in the residual sky subtraction (or a combination of the two).

The simulation data (Figure \ref{xo3b_phase_fake}) show many of the same features, with a few differences.   First, noise on timescales shorter than the wobble period averages quite cleanly to near zero in the binned measurements of centroid and noise pixels, as compared to the same plots for the real dataset.  This suggests the presence in real data of a jitter signal that doesn't integrate to zero in 64 samples, perhaps with a steeper spectrum (more power at low frequencies) than the $1/f$ signal currently included.   Second, the magnitudes of short and long term pointing drift and the amplitude of pointing fluctuations are all larger than in the real data, as seen in the ($x,y$) centroids and $\tilde{\beta}$.  This is also visible in Fig. \ref{xo3b_pos_fake} when compared with Fig. \ref{xo3b_pos_real}.  Third, the simulated backgrounds are much more uniform from one AOR to another because we commanded the same linear increase with time, with constant mean and no offsets between epochs.  Fourth, the larger spread in position has increased the overall noise in the light curve.  This will have consequences for the decorrelation of the measurements and the estimation of eclipse depths.

We display in Figures \ref{lightcurves_real} and \ref{lightcurves_sim} light curves decorrelated using different techniques, for the 5th epochs of the real and simulated observations.  

\begin{figure*}
\plotone{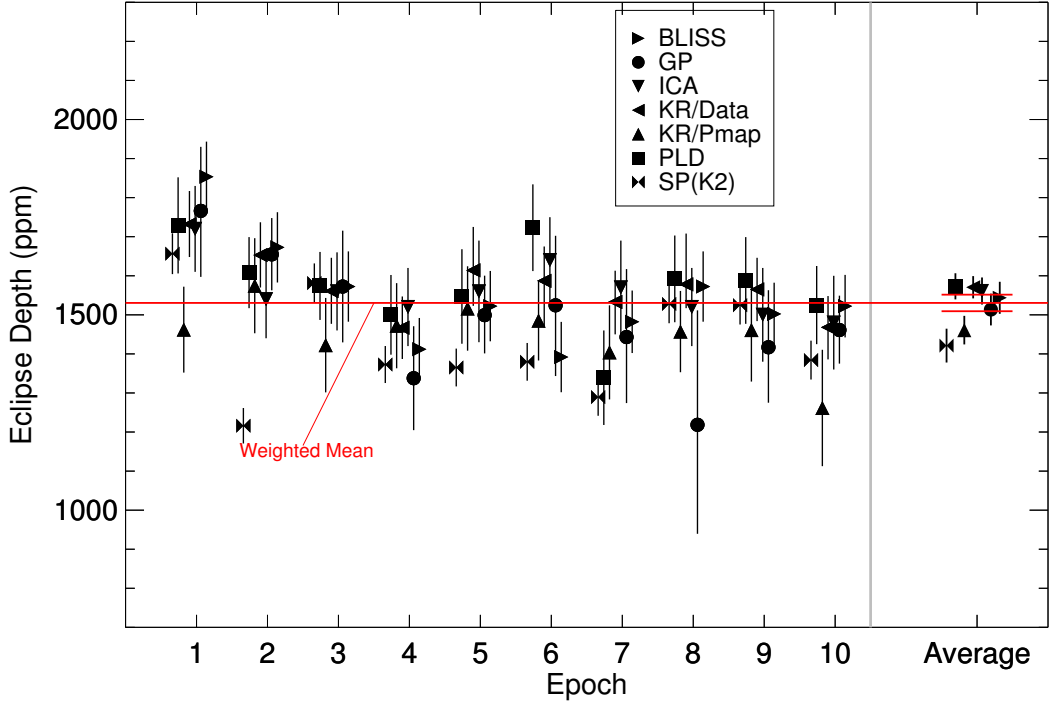}
\caption{Eclipse depths for 10 {\bf real} visits to XO-3b, as computed via various methods.  The group of points for each epoch is separated to minimize confusion.  Error bars in this plot are symmetric; in cases where the technique returned asymmetric uncertainties, we used the largest of the two values. We show the results for the separate visits to the left of the gray vertical line, and the average results to the right. Error bars on the separate visits are the uncertainties reported by the technique.  Error bars on the averages are the uncertainties in the weighted mean, adjusted for ``underdispersion'' by a factor $f_{\litl dis}$ (see text).  The horizontal red lines display the grand mean for all results, $\pm$ its uncertainty.
\label{edepth_real}}
\end{figure*}

\subsection{Eclipse Depths}
\begin{figure*}
\plotone{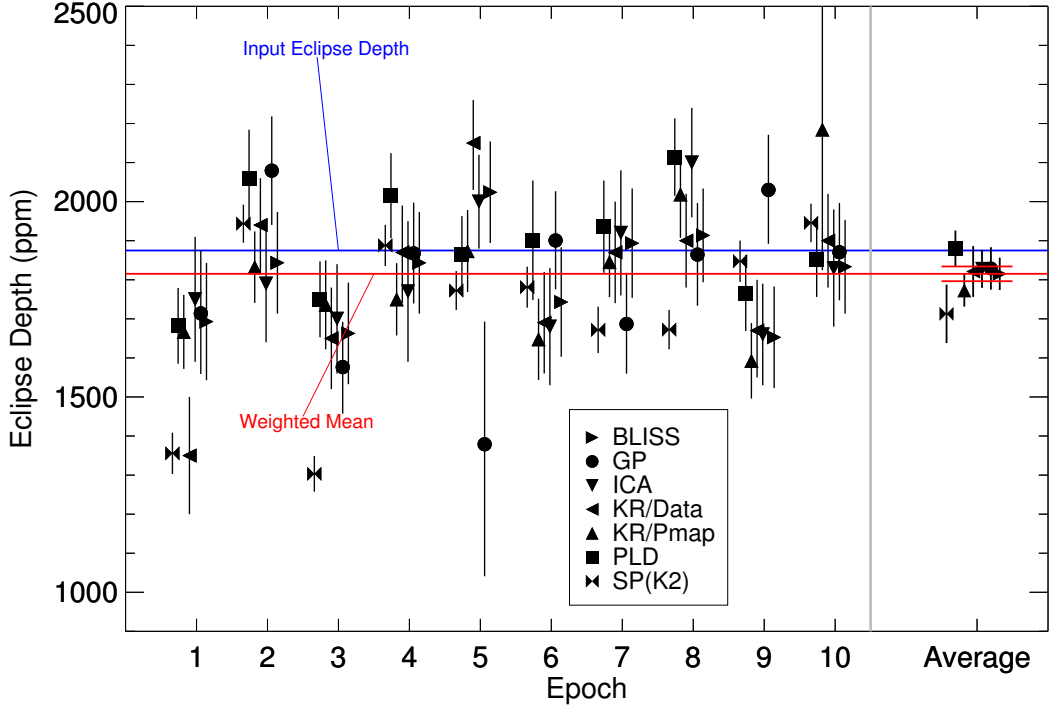}
\caption{Eclipse depths for 10 {\bf simulated} visits to XO-3b, as computed via various methods.  A blue horizontal line indicates the eclipse depth input to the simulations, 1875\,ppm.  See caption for Figure \ref{edepth_real} for further description. 
\label{edepth_fake}}
\end{figure*}

All of the 7 Data Challenge participants estimated eclipse depths and uncertainties from decorrelated light curves, for each set of 10 epochs from the real and simulated datasets.  Figure \ref{edepth_real} plots the measured depths for the real data and Figure \ref{edepth_fake} plots the results for the simulated data.  We define the eclipse depth in terms of the stellar flux: 
\begin{equation}
D = \frac{F_{\litl out}-F_{\litl in}}{F_{\litl in}},
\label{eclipse_depth}
\end{equation}
where $F_{\litl in}$ is the average photometric flux in eclipse (i.e., the stellar flux) and $F_{\litl out}$ is the flux out of eclipse, interpolated to the center of occultation.  
We plot weighted average eclipse depths, $\overbar{D}$, for each of the 7 data reduction methods on the right hand side of Figures \ref{edepth_real} and \ref{edepth_fake}.  The averages are weighted sums of the individual eclipse measurements:
\begin{equation}
\overbar{D}  = \frac{\sum_{i=1}^{N} w_i\,D_i}{\sum_{i=1}^{N} w_i};
\label{weighted_mean}
\end{equation}
where the weights consist of the usual inverse variances, but multiplied by an ``overdispersion'' factor \citep[see][]{1992JPhA...25.1967L}:
\begin{equation}
w_i = \frac{1}{\sigma_i^2\,f_{\litl dis}^2}.
\end{equation}
The factor $f_{\litl dis}$ allows for the possible underestimation of the individual uncertainties, using the scatter in the group of measurements as an additional constraint. We derive it using the $\chi^2$ equation for the mean value (assuming the $D_i$ values are distributed normally about $\overbar{D}$):
\begin{equation}
\chi^2 = \sum_{i=1}^N\frac{\left(D_i - \overbar{D}\right)^2}{\sigma_i^2\,f_{\litl dis}^2} = N-1.
\label{chisquared}
\end{equation}
This can be inverted to solve for $f_{\litl dis}$ (note that, since Equation \ref{weighted_mean} contains $f_{\litl dis}^{-2}$ in both the numerator and denominator, $\overbar{D}$ does not depend on $f_{\litl dis}$):
\begin{equation}
f_{\litl dis}^2 = \sum_{i=1}^N\frac{\left(D_i - \overbar{D}\right)^2}{\sigma_i^2\,(N-1)}.
\label{f_dis}
\end{equation}
\begin{deluxetable*}{lcrrclcrcl}
\tablecaption{Eclipse Depth Statistics: Real Data ($\sigma_{\litl phot}\approx 53\,$ppm)\label{eclipse_depth_stats}}
\tablewidth{0pt}
\tablehead{
\colhead{Method} & \colhead{$\overbar{D}$\tablenotemark{a}} & \colhead{$\overbar{\sigma}$\tablenotemark{b}} & \colhead{SD\tablenotemark{c}} & \colhead{$\sigma_{\litl orig}$\tablenotemark{d}} & \colhead{$f_{\litl dis}$\tablenotemark{e}} & \colhead{$\sigma_{\litl TOT}$\tablenotemark{f}} & \colhead{$R$\tablenotemark{g}} &\colhead{$r$\tablenotemark{h}}& \colhead{Closest Match\tablenotemark{i}} \\
&\colhead{(ppm)}&\colhead{(ppm)}&\colhead{(ppm)}&\colhead{(ppm)}&&\colhead{(ppm)}&\colhead{(ppm)}&&
}\colnumbers
\startdata
BLISS&1543&85&133&27&1.5\ppm{0.5}{0.3}&40&189\phantom{0}&0.40&KR/Data: (--25$\pm$86)\\
GP&1513&152&155&40&1.0&40&220\phantom{0}&0.34&BLISS: (--60$\pm$121)\\
ICA&1560&111&71&34&1.0&34&101\phantom{0}&0.74&KR/Data: (--14$\pm$56)\\
KR/Data&1570&94&79&28&1.0&28&113\phantom{0}&0.66&ICA: (14$\pm$56)\\
KR/Pmap&1460&117&81&36&1.0&36&116\phantom{0}&0.65&SP(K2): (21$\pm$172)\\
PLD&1573&107&111&33&1.0&33&158\phantom{0}&0.48&KR/Data: (--3$\pm$86)\\
SP(K2)&1421&48&137&15&2.8\ppm{1.0}{0.5}&43&195\phantom{0}&0.39&KR/Pmap: (--21$\pm$172)\\
\hline
Average\tablenotemark{j} &1520&102&110&30&1.3&36&156\phantom{0}&0.52&\multicolumn{1}{c}{\nodata}\\
\enddata
\tablenotetext{a}{Weighted mean eclipse depth over the 10 AOR measurements of XO-3b.}
\tablenotetext{b}{Mean eclipse depth uncertainty reported for the 10 AOR measurements.}
\tablenotetext{c}{Sample standard deviation in eclipse depth over the 10 AORs.}
\tablenotetext{d}{Weighted uncertainty in the mean eclipse depth, based only on the originally reported uncertainties.}
\tablenotetext{e}{``Dispersion factor'' that multiplies the uncertainties, required to make $\chi_\nu^2 = 1$ (see text).}
\tablenotetext{f}{Total uncertainty in the mean, after being corrected for dispersion, $\sigma_{\litl TOT} = f_{\litl dis}\,\sigma_{\litl orig}$.}
\tablenotetext{g}{The ``repeatability,'' ie., the standard deviation in differences between pairs of eclipse depth measurements.}
\tablenotetext{h}{The ``reliability'' of the technique, $\sigma_{\litl phot}/{\rm SD}$.}
\tablenotetext{i}{Technique with the closest range in eclipse values to this one, followed by(Mean$\pm$SD) difference.}
\tablenotetext{j}{Straight averages along the columns.}
\end{deluxetable*}

\begin{deluxetable*}{lcrrclcrclcrc}
\tablecaption{Eclipse Depth Statistics: Simulated Data ($\sigma_{\litl phot}\approx 53\,$ppm)\label{eclipse_depth_stats_sim}}
\tablewidth{0pt}
\tablehead{
\colhead{Method} & \colhead{$\overbar{D}$} & \colhead{$\overbar{\sigma}$} & \colhead{SD} & \colhead{$\sigma_{\litl orig}$} & \colhead{$f_{\litl dis}$} & \colhead{$\sigma_{\litl TOT}$} & \colhead{$R$} &\colhead{$r$}& \colhead{Closest Match} &\colhead{RMSE\tablenotemark{a}} & \colhead{$\overbar{B}$\tablenotemark{b}} & \colhead{$a$\tablenotemark{c}}\\
&\colhead{(ppm)}&\colhead{(ppm)}&\colhead{(ppm)}&\colhead{(ppm)}&&\colhead{(ppm)}&\colhead{(ppm)}&&&\colhead{(ppm)} & \colhead{(ppm)}&
}\colnumbers
\startdata
BLISS&1815&131&120&41&1.0&41&171\phantom{0}&0.44&ICA: (--9$\pm$76)&131&-59\phantom{00}&0.40\\
GP&1829&154&211&43&1.2\ppm{0.4}{0.2}&54&300\phantom{0}&0.25&BLISS: (--13$\pm$277)&215&-45\phantom{00}&0.25\\
ICA&1827&148&144&45&1.1\ppm{0.4}{0.2}&48&205\phantom{0}&0.37&BLISS: (9$\pm$76)&148&-47\phantom{00}&0.36\\
KR/Data&1821&128&217&40&1.6\ppm{0.6}{0.3}&65&309\phantom{0}&0.24&BLISS: (--11$\pm$129)&219&-53\phantom{00}&0.24\\
KR/Pmap&1772&125&180&32&1.3\ppm{0.4}{0.2}&41&256\phantom{0}&0.29&ICA: (--5$\pm$137)&181&-102\phantom{00}&0.29\\
PLD&1880&108&140&33&1.4\ppm{0.5}{0.2}&45&199\phantom{0}&0.38&BLISS: (83$\pm$114)&134&5\phantom{00}&0.39\\
SP(K2)&1712&51&226&16&4.6\ppm{1.6}{0.8}&74&322\phantom{0}&0.23&KR/Data: (--80$\pm$191)&266&-162\phantom{00}&0.20\\
\hline
Average &1808&121&177&36&1.7&53&252\phantom{0}&0.32&\multicolumn{1}{c}{\nodata}&185&-66\phantom{00}&0.30\\
\enddata
\tablenotetext{a}{Root mean square deviation of the 10 individual measurements from the input eclipse depth of 1875 ppm.}
\tablenotetext{b}{The mean bias, or deviation of $\overbar{D}$ from the input eclipse depth.}
\tablenotetext{c}{Accuracy of technique, $\sigma_{\litl phot}/$RMSE.}
\end{deluxetable*}


The total variance in the mean is given by the inverse sum of weights:
\begin{eqnarray}
\sigma_{\litl TOT}^2 & = & \frac{1}{\sum_{i=1}^N w_i} \nonumber\\
                     & = & \frac{1}{\sum_{i=1}^N\frac{1}{\sigma_i^2\,f_{\litl dis}^2}} \nonumber \\
                     & = & f_{\litl dis}^2 \, \sigma_{\litl orig}^2,
\end{eqnarray}                     
where $\sigma_{\litl orig}$ is the original uncertainty in the mean derived from $w_i = 1/\sigma_i^2$ (i.e., $f_{\litl dis}=1$).  

In Tables \ref{eclipse_depth_stats} (real) and \ref{eclipse_depth_stats_sim} (simulated) we list the values for $\overbar{D}$, the mean uncertainty $\overbar{\sigma}$, the standard deviation in depth, (SD), as well as $\sigma_{\litl orig}$, $f_{\litl dis}$, and $\sigma_{\litl TOT}$ for each technique.  Wherever ${\rm SD} \gtrsim \overbar{\sigma}$, one expects that the uncertainties have been underestimated, and indeed in all instances where this holds, $f_{\litl dis} > 1$.  For the real data, only two techniques had underestimated uncertainties ($f_{\litl dis} > 1$), and for both real and simulated data only one technique, SP(K2), which was not developed for \Spitzer\ data, has $f_{\litl dis} > 2$.

Since the sum in Equation \ref{chisquared} defines a $\chi^2$ probability distribution with $N-1$ degrees of freedom, we derive a 68\% confidence interval on $f_{\litl dis}$ as those values for which the distribution obtains 16\% and 84\% of its integrated area.  The resulting intervals are specified on Tables \ref{eclipse_depth_stats} and \ref{eclipse_depth_stats_sim} as positive and negative error bars on $f_{\litl dis}$.  

\subsection{The photon limit}
Because our goal in this paper is to assess the potential variability in eclipse depth measurements we must first calculate the noise floor for the real and simulated datasets, i.e., the intrinsic variability due to photoelectron counting statistics and readout noise.  

We estimate the signal-to-noise ratio for a single 2\,s data frame based on aperture photometry.  Combining Equations 5 and 13 of \citet{Garnett:1993vi}, the variance in Fowler-sampled electron counts (including the effects of readout noise) is
\begin{equation}
\sigma^2_e = \sigma_{\litl int}^2 \,{
\left[\frac{2\sigma_{\litl rn}^2}{\sigma_{\litl int}^2\cdot({\rm FN})} + 1 - \frac{2({\rm FN})}{3n_{\litl max}} + \frac{1}{6({\rm FN})\cdot n_{\litl max}}\right]}
\label{gar_for_var}
\end{equation}
where $\sigma_{\litl int}^2$ is the equivalent shot noise variance in electron counts accumulated over the {\it integration} time ($t_{\litl int} = n_{\litl max}\Delta t$), FN is the Fowler number, $\sigma_{\litl rn}$ is the standard deviation of the readout noise (per read), $n_{\litl max} = 2({\rm FN}) + {\rm WT}$ is the total number of Fowler samples per integration, WT is the number of wait ticks, and $\Delta t$ is the sample time (see Section \ref{fowler_section} for more information on Fowler sampling with IRAC).  For 2\,s subarray measurements, FN=8, WT=184, $\Delta t = 0.01\,$s, and $\sigma_{\litl rn} = 9.4\,$e. The shot noise variance has the same value as the total electron counts accumulated over the entire integration, $\sigma_{\litl int}^2 \approx F_e \,(t_{\litl int}/t_{\litl exp})$.  [The scale factor $t_{\litl int}/t_{\litl exp}$ is necessary because Fowler sampling returns $F_e$, the accumulated charge per {\it exposure} time, $t_{\litl exp} = ({\rm FN + WT})\Delta t$].  

To estimate $\sigma_e$ we average over the entire multi-epoch photometric dataset of XO-3b to obtain values for $\overbar{F}_{\litl ap-bg}$, the  number of electrons measured in the source aperture after background subtraction (i.e., the signal); $\overbar{F}_{\litl ap}$, the number of electrons in the aperture {\it before} background subtraction (from which we derive the noise in the aperture); and $\overbar{F}_{\litl bg}$, the  number of electrons in the background annulus (from which we derive the noise in the background).  For real data, we obtain $\overbar{F}_{\litl ap-bg}=70858\,$e, $\overbar{F}_{\litl ap}=70917\,$e, and $\overbar{F}_{\litl bg}=463\,$e; for the simulations, $\overbar{F}_{\litl ap-bg}=73246\,$e, $\overbar{F}_{\litl ap}=73358\,$e, and $\overbar{F}_{\litl bg}=881\,$e.    

The first term in square brackets of Equation \ref{gar_for_var}, when divided by the remaining three terms, gives the relative contribution of readout noise to $\sigma^2_e$.  For our integration parameters, this term equals $21.9/F_e$, which is much less than $1-2{\rm (FN)}/(3n_{\litl max}) + 1/[6{\rm (FN)}\cdot n_{\litl max}]= 0.98$ if $F_e \gg 22.3\,$e.  Thus, readnoise is insignificant for XO-3b, where $F_e\sim 70000\,$e.

Substituting $\overbar{F}_{\litl ap}$ and $\overbar{F}_{\litl bg}$ into Equation \ref{gar_for_var} [using $\sigma_{\litl int}^2 = F \,(t_{\litl int}/t_{\litl exp})$] yields noise variances for the aperture and background, $\sigma^2_{\litl ap}$ and $\sigma^2_{\litl bg}$.  Their sum equals the noise variance for an aperture photometry measurement: $\sigma^2_{\litl ap-bg} = \sigma^2_{\litl ap}+\sigma^2_{\litl bg}$.  We obtain $\sigma_{\litl ap-bg} = 268\,$e for both real and simulated data.  Dividing these into $\overbar{F}_{\litl ap-bg}$ gives the expected signal-to-noise ratios for a single photometric data point:  ${\rm (S/N)}_{\litl single}^{\litl real} = 264$ and ${\rm (S/N)}_{\litl single}^{\litl sim} = 268$.  These numbers are extremely close to the square roots of the background-subtracted aperture fluxes, which means that neither the backgrounds nor readout noise are significant determinants of S/N for XO-3b.  From this point on, we refer to the intrinsic variability as {\it photon} noise.

We now propagate the expected photon noise error in a single photometric measurement to that for the entire eclipse depth measurement.  Recall Equation \ref{eclipse_depth} for the eclipse depth, which can be rewritten:
\begin{equation}
D = \frac{F_{\litl out}}{F_{\litl in}} - 1.
\end{equation}
The photon noise variance in the eclipse depth is the variance in $F_{\litl out}/F_{\litl in}$:
\begin{equation}
\sigma_{\litl phot}^2 = (1+D)^2\left[\left(\frac{\sigma_{\litl out}}{F_{\litl out}}\right)^2 +
                                     \left(\frac{\sigma_{\litl in}}{F_{\litl in}}\right)^2\right].
\end{equation} 
Since $F_{\litl out}$ and $F_{\litl in}$ are the average fluxes inside and outside eclipse, we have:
\begin{equation}
\left(\frac{\sigma_{\litl in}}{F_{\litl in}}\right)^2 = \frac{1}{N_{\litl in}}\, \left(\frac{\sigma_{\litl single}}{F_{\litl single}}\right)^2,
\end{equation} 
and similarly for the out-of-eclipse flux.  We define $N_{\litl in}$ and $N_{\litl out}$ as the total number of frames in and out of eclipse.  Let  $N_{\litl in} = f_{\litl in} N$, where the total number of measured frames is $N=14912$ (real) and 15232 (simulated).  Keep in mind that the flux outside of eclipse, $F_{\litl out}$, is a factor $D+1$ larger than $F_{\litl in}$.  Also, substitute $(\sigma_{\litl single}/F_{\litl single})^2 = 1/(S/N)_{\litl single}^2$.  

The photon noise variance in the eclipse depth consequently becomes:
\begin{equation}
\sigma_{\litl phot}^2 = \frac{(1+D)^2}{N\,(S/N)_{\litl single}^2} \left[\frac{1}{(1-f_{\litl in})(1+D)} 
+ \frac{1}{f_{\litl in}}\right].
\label{photon_noise}
\end{equation}
If we use $f_{\litl in} = 1/3$ and assume eclipse depths of $D_{\litl real} = 1520\,$ppm (average measured value) and $D_{\litl sim} = 1875\,$ppm (actual input value), we find that the expected variability in the eclipse depth due to photon noise is $\sigma_{\litl phot} = 53\,$ppm, for both real and simulated data. 
 
\begin{figure*}
\plotone{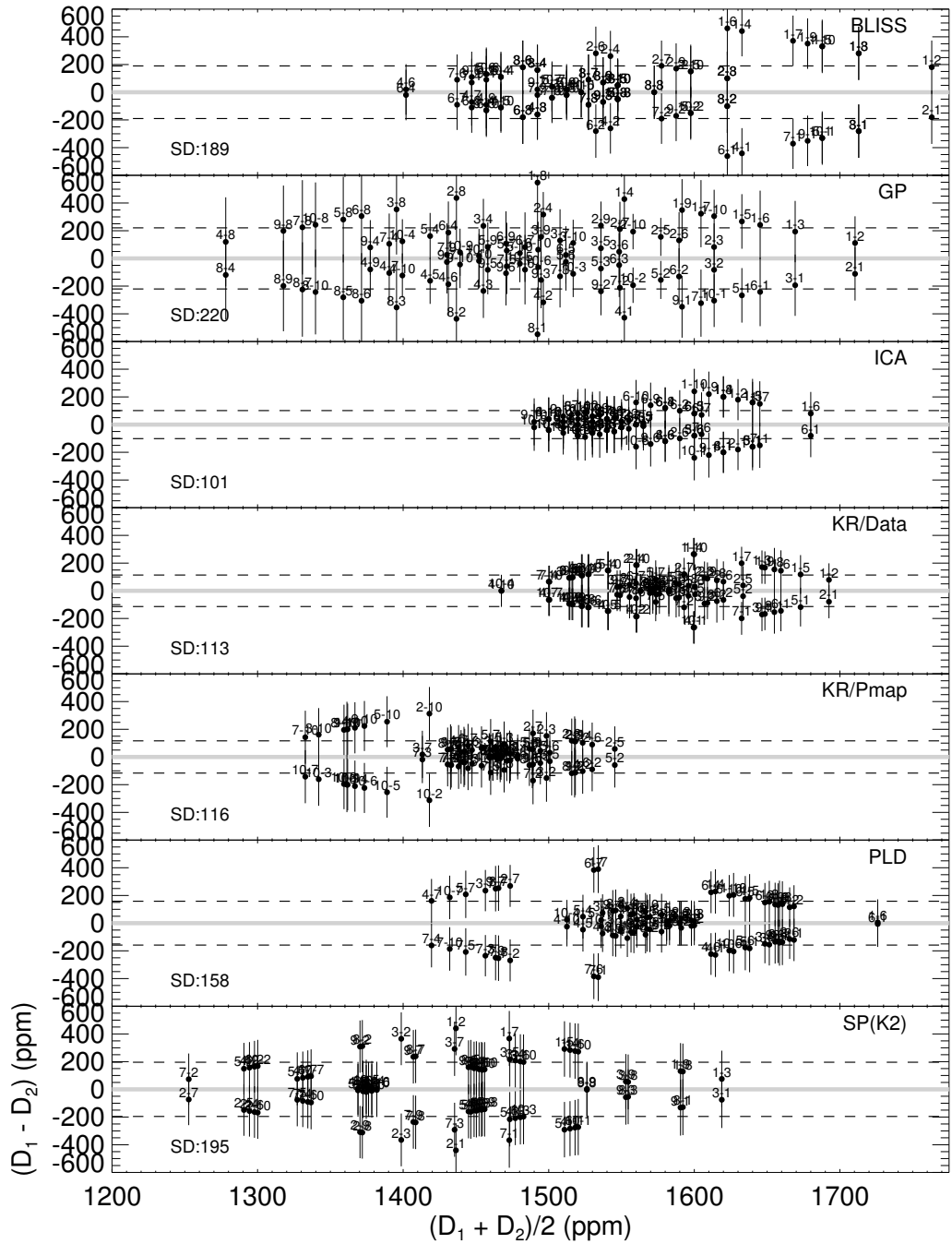}
\caption{Mean/difference plots for repeated visits to XO-3b, {\bf real} data.  Each panel shows the difference between all pairs of eclipse depth measurements for a given reduction method, as a function of the average of the two depths.  Each point is labeled with the two epochs being compared.  Two horizontal dashed lines indicate $\pm$ one standard deviation of the differences (repeatability), also labeled in the lower left corner of the panel.  A gray line indicates $(D_1-D_2)=0$. The horizontal spread of the data relates to the precision of the set of measured depths, whereas the vertical spread indicates their repeatability.  
\label{diff_real_visits}}
\end{figure*}
\begin{figure*}
\plotone{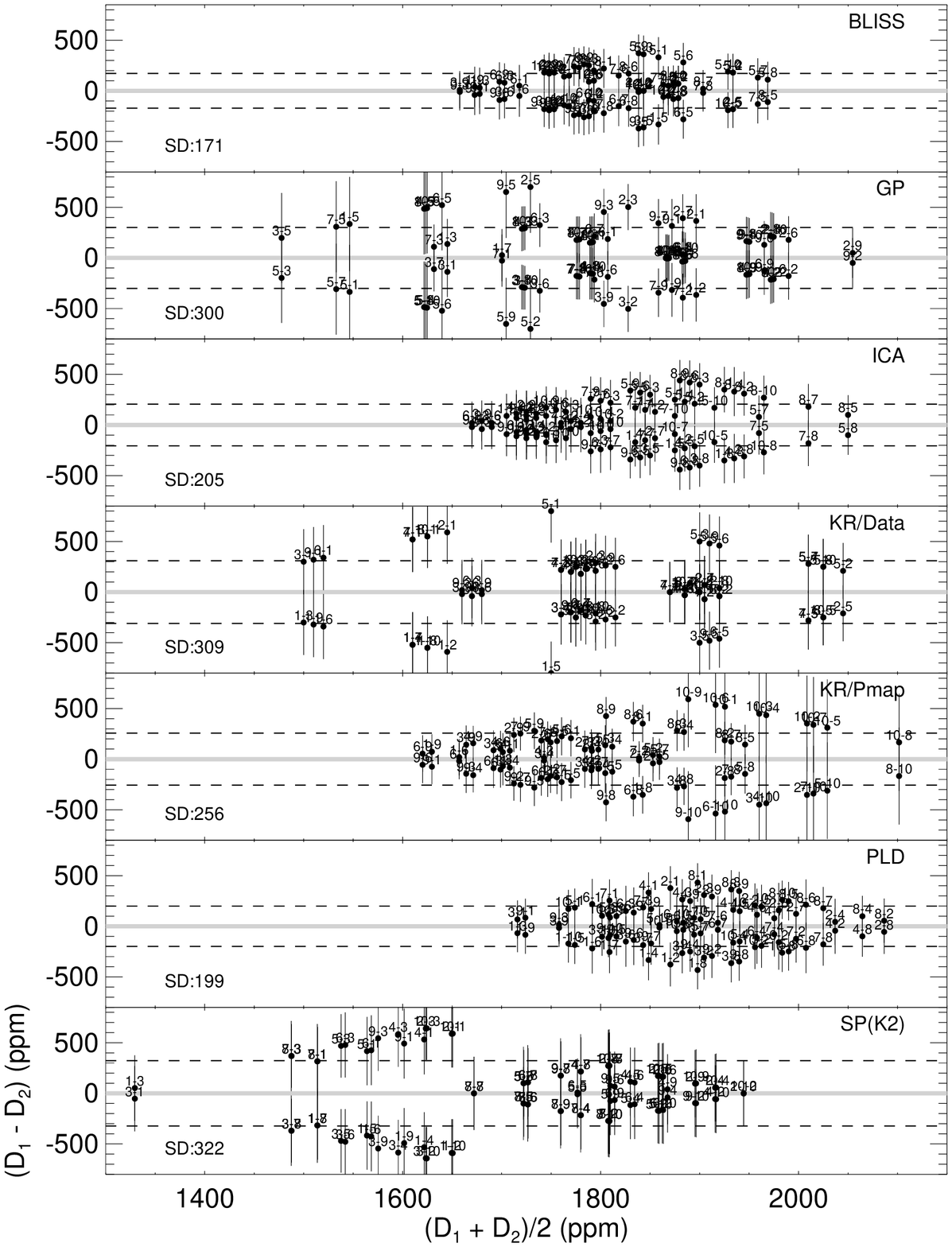}
\caption{Mean/difference plots for repeated visits to XO-3b, {\bf simulated} data.  See caption to Figure \ref{diff_real_visits} for further description. 
\label{diff_fake_visits}}
\end{figure*}

\subsection{Repeatability, reliability, and accuracy}
A substantial literature exists in other scientific fields discussing techniques for estimating the repeatability, reliability, and accuracy of a set of measurements \citep[see, for example,][for discussions of repeatability and reliability]{Altman:1983bn,Bartlett:2008bc}.  We review and adapt these terms below.  

\subsubsection{Repeatability}
We define the repeatability, $R$, to be the value below which we can expect the difference between two eclipse depth measurements to lie 68\% of the time, for a given data reduction method.  For our purposes, $R$ equals the standard deviation (SD) of the differences in separate measurements made with the same method, $R = {\rm SD}(\Delta_{ij})$.  Repeatability has the same units as the measurements themselves (e.g., ppm).  Note that the repeatability is not the standard deviation of the measurements, which indicates the spread in depths around the mean value, but of their differences.  

One way to assess repeatability visually is with ``mean/difference'' plots \citep{Altman:1983bn}, which we show in Figures \ref{diff_real_visits} (real) and \ref{diff_fake_visits} (simulated).  The plots display, for all pairs of measured eclipses, the difference in depth ($\Delta_{ij}$) as a function of the pair average eclipse depth.  (To obtain a statistically valid estimate for this comparison, each pair must be counted twice, with the order of the indices reversed.)  Mean/difference plots often show more clearly the limits of variability of the difference between sets than, for example, correlation plots where the variables are plotted against each other. In mean/difference plots, the horizontal spread of the data (spread in average values in paired epochs) is related to the precision of the measurements (when the overall scatter in values is large, the midpoint between pairs of values will have a relatively large spread). The vertical spread in mean/difference plots indicates the repeatability, i.e., how far apart we expect two separate measurements to be.  Specifically, we compute $R$ from the standard deviation of each group of paired differences, labeled ``SD'' on the bottom left of each frame of Figures \ref{diff_real_visits} and \ref{diff_fake_visits}.

Patterns in mean/difference plots can sometimes elucidate patterns in the data, but they need to be examined carefully, because of the inherent correlation between the data axes.  If we define $x \equiv (D_1+D_2)/2$ (the horizontal axis) and $y \equiv (D_1-D_2)$ (the vertical axis), it is apparent that the two axes are not independent: the relationship between $y$ and $x$ can be written either $y = 2(D_1-x)$ or $y = 2(x + D_2)$.  Thus, for a given $D_1$ or $D_2$, the inter-epoch difference ($y$) is expected to follow a linear trend as a function of the inter-epoch average ($x$).  This trend is indeed visible in Figures \ref{diff_real_visits} and \ref{diff_fake_visits} if we group by epoch.  It is most visible when either (1) $D_1$ or $D_2$ is significantly different from the average depth, or (2) the inter-epoch average, $x$, has a large spread.  For example, in the real data four of the methods (BLISS, GP, ICA, KR/Data, and PLD) show  an inverse linear relationship between $x$ and $y$ for paired differences involving epoch 1 (labeled ``1-2'', ``1-3'', etc.).  This is because the epoch 1 depth is systematically high for each of these methods (as one might also guess from Fig. \ref{edepth_real}).  

The values of $R$ for each technique are listed in column 8 of Tables \ref{eclipse_depth_stats} and \ref{eclipse_depth_stats_sim}.  The real XO-3b results show a repeatability of better than 220\,ppm in all cases, with an average value of $\overbar{R} = 156\,$ppm.  The simulations are less repeatable, with $\overbar{R} = 252\,$ppm.  This is probably due to the presence of more noise in the eclipse depth measurements for the simulations, as expected from the greater pointing scatter (\S3.1). To confirm that the repeatability as computed is consistent with our definition above, we have constructed cumulative distributions of each set of eclipse depth differences.  As expected, most measured differences are less than $R$ 68\% of the time, for both real and simulated data.  In only a few cases, the 68th percentile is as much as 20\,ppm larger than $R$.  

Strictly speaking, in earthbound experiments repeatability is usually assessed on consecutive measurements under {\it identical} conditions.\footnote{The measurement of differences under changing conditions is often called {\it reproducibility}.}    This is not possible for eclipses, since they cannot be repeated at will.  In the time between eclipses, the experimental situation will likely change: a new pointing center and different pointing jitter can change the correlated noise properties; exposure of the detector arrays to other sources of photons may produce latent charge on the pixels of interest, or existing latent charge may decay; the planetary phase curve and eclipse timing and depth may not be the same from one orbit to the next due to stellar variability, perturbations of the planet's orbit, or atmospheric evolution.  But for consistency with the astrophysics literature, we will continue to refer to the spread in eclipse depth differences as repeatability.

\begin{figure}
\plotone{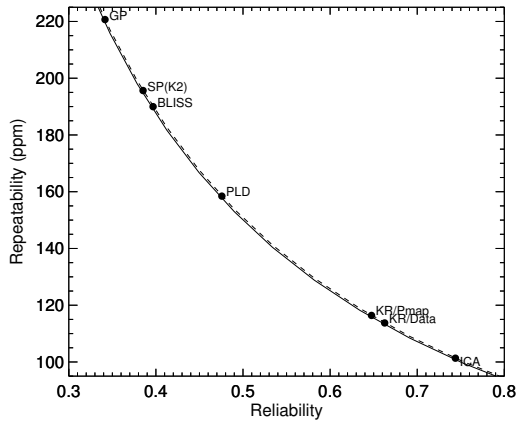}
\caption{Repeatability as a function of reliability, for the {\bf real} XO-3b eclipse depth measurements. The dashed curve displays the fit $R=75.4\,r^{-1}\,$ppm, and the solid curve shows the expected behavior $R=\sqrt{2}\sigma_{\litl phot}r^{-1}=75\,r^{-1}$.
\label{real_rep_rel}}
\end{figure}
\begin{figure}
\plotone{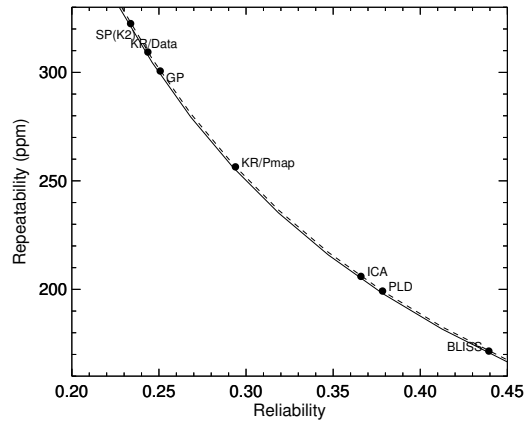}
\caption{Repeatability as a function of reliability, for the {\bf simulated} data.  The dashed curve displays the fit $R=(75.4\,r^{-1})\,$ppm, and the solid curve shows the theoretical behavior, $R=75\,r^{-1}$. 
\label{fake_rep_rel}}
\end{figure}
\begin{figure}
\plotone{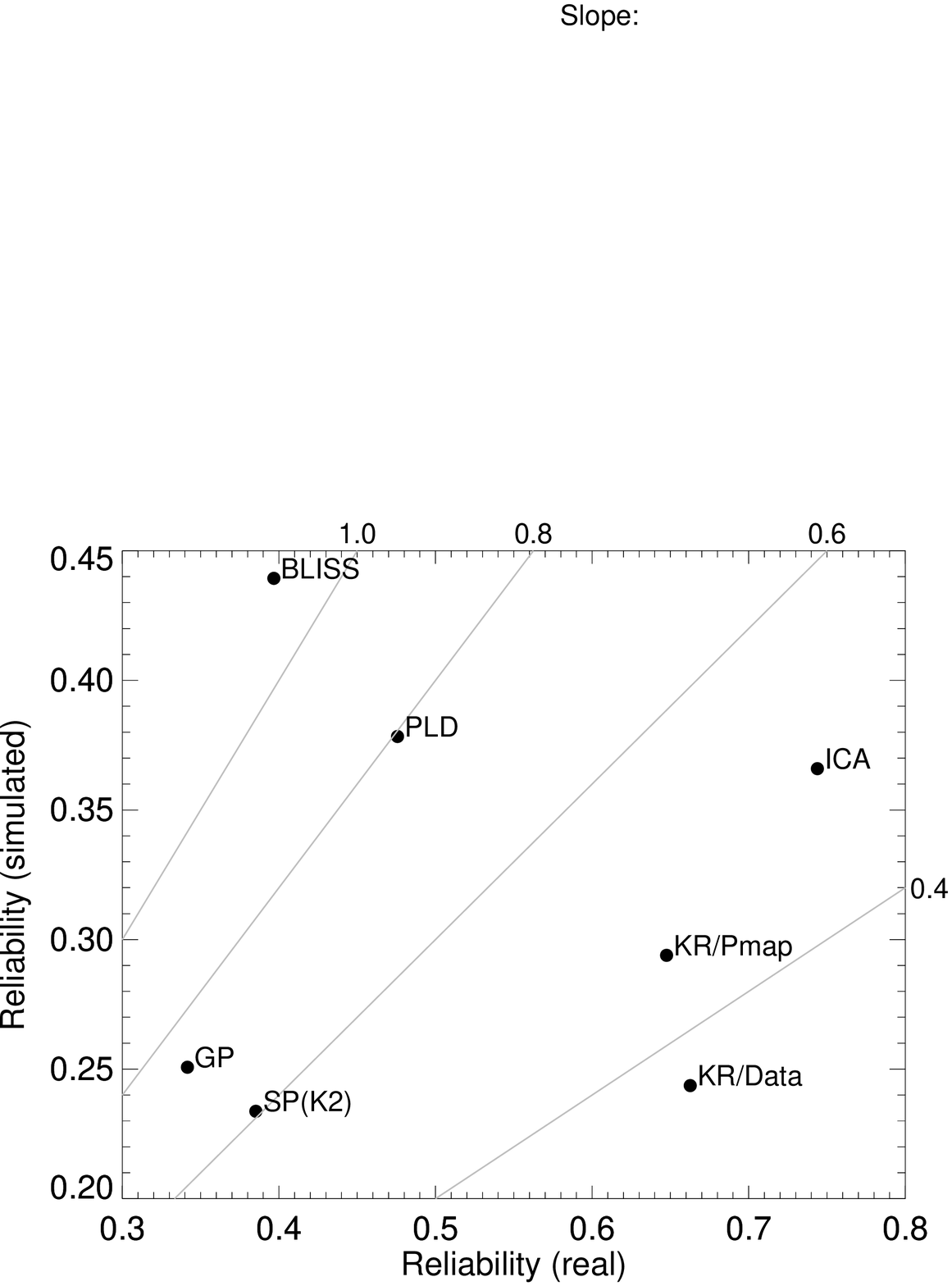}
\caption{Reliability comparison between simulated and real eclipse depths.  Gray lines indicate $r_{\litl sim}/r_{\litl real} = $0.4, 0.6, 0.8, and 1.0.  
\label{rel_real_fake}}
\end{figure}

\subsubsection{Reliability}
We define the {\it reliability}, $r$, to be the ratio between the intrinsic variability of a set of measurements (in the absence of astrophysical variation) to their observed variability, for a given method. In the context of eclipse depth measurements, the intrinsic variability is the standard deviation in the depth due only to photon noise, $\sigma_{\litl phot}$ (Equation \ref{photon_noise}), and the observed variability is the measured standard deviation in the depth, ${\rm SD}$.  The measured variance combines both the photon noise and the variance due to ``measurement error,'' caused by residual correlated noise.  (Here we assume no variability in the planetary system, but its presence would add to the {\it measured} variance and decrease the reliability.)  The value of $r \equiv \sigma_{\litl phot}/({\rm SD})$ is unitless and can range from 0 (all scatter due to measurement error) to 1 (no measurement error).  Reliability is essentially a normalized measure of precision, and is inversely related to repeatability (we demonstrate this relationship below).  

We list the computed values of $r$ for each method in column 9 of Tables \ref{eclipse_depth_stats} and \ref{eclipse_depth_stats_sim}.  For the real data, the reliability is quite high in most cases, with an average of $\overbar{r}=0.52$, suggesting that half of the scatter is due to intrinsic photon noise.  The ICA and kernel regression (KR/Data and KR/Pmap) techniques appear to have the least amounts of correlated noise (scatter in eclipse depths consistent with more than half photon noise).  For the simulated data, however, the values are lower, with an average reliability of $\overbar{r}=0.32$.  

Figures \ref{real_rep_rel} and \ref{fake_rep_rel} are scatterplots of repeatability vs. reliability for the real and simulated eclipse depths, respectively. These data appear inversely correlated, which is not surprising.  If two values are drawn from the same parent population, the variance in the difference between the values should be twice the variance of the original distribution, which means that for large enough samples $[{\rm SD}(\Delta_{ij})]^2 = 2({\rm SD})^2$.  Thus by the definition of $r$, we expect $R = \sqrt{2}\sigma_{\litl phot}r^{-1}$.  We overlay this theoretical curve, as well as linear fits to $R$ as a function of $r^{-1}$ on Figures \ref{real_rep_rel} and \ref{fake_rep_rel}. The two curves for each plot are practically identical, with the fit factors multiplying $r^{-1}$ within 1\% of the theoretical values, indicating statistical self-consistency between $[{\rm SD}(\Delta_{ij})]$ and SD.  This implies that the repeatability and reliability derived from 10-element samples are robust.

Figure \ref{rel_real_fake} plots the reliability for simulated data as a function of that for real data, for the seven decorrelation methods, with lines of different slope overlaid. There seems to be no relationship between the reliability measures for real and simulated eclipses, except that the simulated values are nearly all lower than their real counterparts.   Only BLISS has a similar reliability for both real and simulated data ($r=0.40$ and 0.44, respectively).   The kernel regression techniques both show the largest decrease, with $r_{\litl sim} \approx 0.4\,r_{\litl real}$. We conclude that BLISS is most robust to increases in positional dispersion, the main source of additional correlated noise between the simulated and real datasets.  The (gaussian) kernel regression methods seem to be least robust to such changes.

\subsubsection{Accuracy}

The {\it accuracy} of a technique is a quantitative estimate of how well the technique measures a given characteristic of a system.  Earlier definitions of accuracy were synonymous with what is now called {\it trueness}, the proximity of the {\it mean} of a set of measurements to the true value.  Current definitions of accuracy, however, encompass both random {\it and} systematic error.  That is, accuracy is limited by precision\footnote{ISO 5725-1: 1994, ``Accuracy (trueness and precision) of measurement methods and results.''}.  Even if the mean of a set of measurements is extremely close to the truth (bias is low and trueness is high), if the reliability (precision) is low (the scatter in results is large), the result is still considered to have low accuracy.  

Assume an exoplanet system is observed $N$ times, and a given technique $j$ yields a set of measurements of the eclipse depth, $\{D_{ij}\}$ ($i=1,...,N$), with average value, $\overbar{D_j}$.  Let the true depth be $D_t$.  We can think of a measurement of eclipse depth as being the sum of the true value, any bias in that measurement (systematic error), $B_{ij}$, and two random noise terms:
\begin{equation}
D_{ij} = D_t + B_{ij} + \epsilon_{ij}^{\litl phot} + \epsilon_{ij}^{\litl meas}.
\end{equation}
Here, $\epsilon_{ij}^{\litl phot}$ is the error in measurement $ij$ due to photon noise and $\epsilon_{ij}^{\litl meas}$ is the random measurement error (e.g. a random component of residual correlated noise).  These error terms can be thought of as samples of random variables with means of 0 and standard deviations $\sigma_{\litl phot}$ and $\sigma_{\litl meas}$.  Taking the mean of $D_{ij}$ gives:
\begin{equation}
\overbar{D} = D_t + \overbar{B}.
\end{equation}
Thus the average measured value is approximately the sum of the true value and the average bias.  Alternately, if we know $D_t$ (as we do for the simulations), we can estimate the mean bias as 
\begin{equation}
\overbar{B} = \overbar{D}-D_t.
\end{equation}

The scatter in the data about the true value is measured by the mean square error:
\begin{eqnarray}
{\rm MSE}& = &\frac{1}{N}\sum_{i=1}^N (D_{ij}-D_t)^2.\label{mse}\\
         & \approx &\overbar{(B^2)} + \sigma_{\litl phot}^2 + \sigma_{\litl meas}^2.\nonumber
\end{eqnarray}

We now define accuracy using the square root of MSE, analagous to using SD for reliability:
\begin{equation} 
a \equiv \sigma_{\litl phot}/{\rm RMSE}.
\end{equation}
This has the desired limiting behavior:  if the bias is minimized ($\overbar{B}\rightarrow 0$; $\overbar{D} \rightarrow D_t$), MSE approaches $\sigma_{\litl phot}^2 + \sigma_{\litl meas}^2 = ({\rm SD})^2$ and the accuracy approaches the reliability; but as the bias increases, $a \rightarrow 0$.  

Columns 11--13 of Table \ref{eclipse_depth_stats_sim} list the root mean square error (RMSE), the average bias, $\overbar{B}$, and the accuracy of each technique applied to the simulated XO-3b eclipses.    
\begin{figure}
\plotone{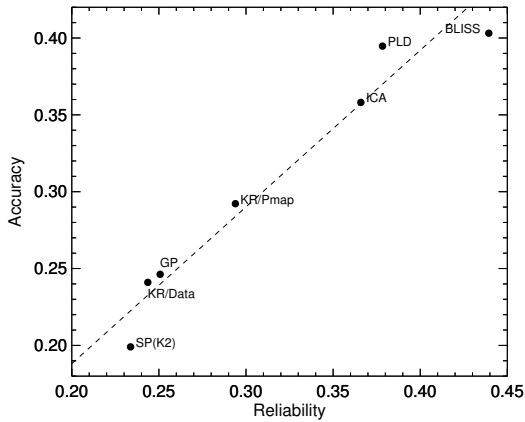}
\caption{Accuracy vs. reliability, as defined in the text, for the simulated eclipse depth measurements.  It is better to be on the upper right of the plot (lower scatter, lower error) than on the lower left.  The dashed line displays the fit $a = 1.02\,r -0.02$, which confirms that on average, the techniques have minimal bias.
\label{edepth_sim_v_r}}
\end{figure}

Figure \ref{edepth_sim_v_r} plots $a$ as a function of $r$.  This figure shows how well a technique (1) can be relied on to give the {\it same} eclipse depth over multiple epochs where the true depth is constant (reliability: ratio of intrinsic to measured scatter: bottom axis); and (2) can be expected to give the {\it correct} eclipse depth over multiple epochs (accuracy: ratio of intrinsic to measured error: left axis).   It is better to be on the upper right of the plot (lower scatter, lower error) than on the lower left.

The majority of the methods have RMSE values similar to their SD values, and thus accuracy nearly equal to reliability.  We plot a fit to the data in Fig. \ref{edepth_sim_v_r}, $a = 1.02\,r-0.02$, which confirms that {\it on average} the limiting value $a\approx r$ is reached for these techniques.  In other words, the bias is within one standard deviation of zero. 

\begin{figure}
\plotone{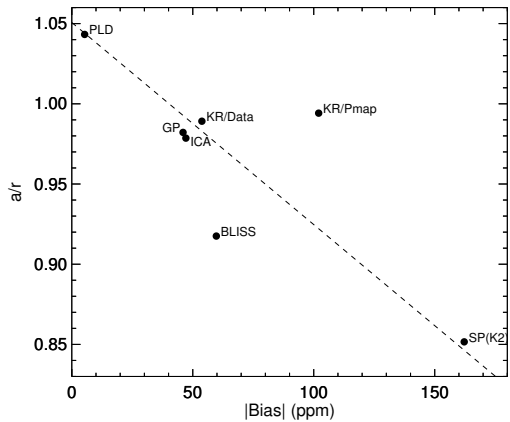}
\caption{The accuracy/reliability ratio as a function of mean absolute bias for the simulated eclipse depth measurements.  The dashed line displays the fit $a/r = 1.1-0.0013\,|\overbar{B}|$.
\label{edepth_sim_v_r_bias}}
\end{figure}

In detail the ratio $a/r$, which equals $({\rm SD})/{\rm RMSE}$, is not unity but varies by 20\% among the techniques.  We display $a/r$ as a function of mean absolute bias in Figure \ref{edepth_sim_v_r_bias}.  The ratio is (roughly) inversely proportional to $|B|$.  This can be understood theoretically if we write:
\begin{eqnarray}
(a/r)^2 & \approx &\frac{\sigma_{\litl phot}^2 + \sigma_{\litl meas}^2}{\overbar{(B^2)} + \sigma_{\litl phot}^2 + \sigma_{\litl meas}^2}\\
    & = & \left[\frac{\overbar{(B^2)}}{({\rm SD})^2} + 1\right]^{-1}.
\end{eqnarray}

\subsection{Comparison Between Methods}
\begin{figure*}
\plotone{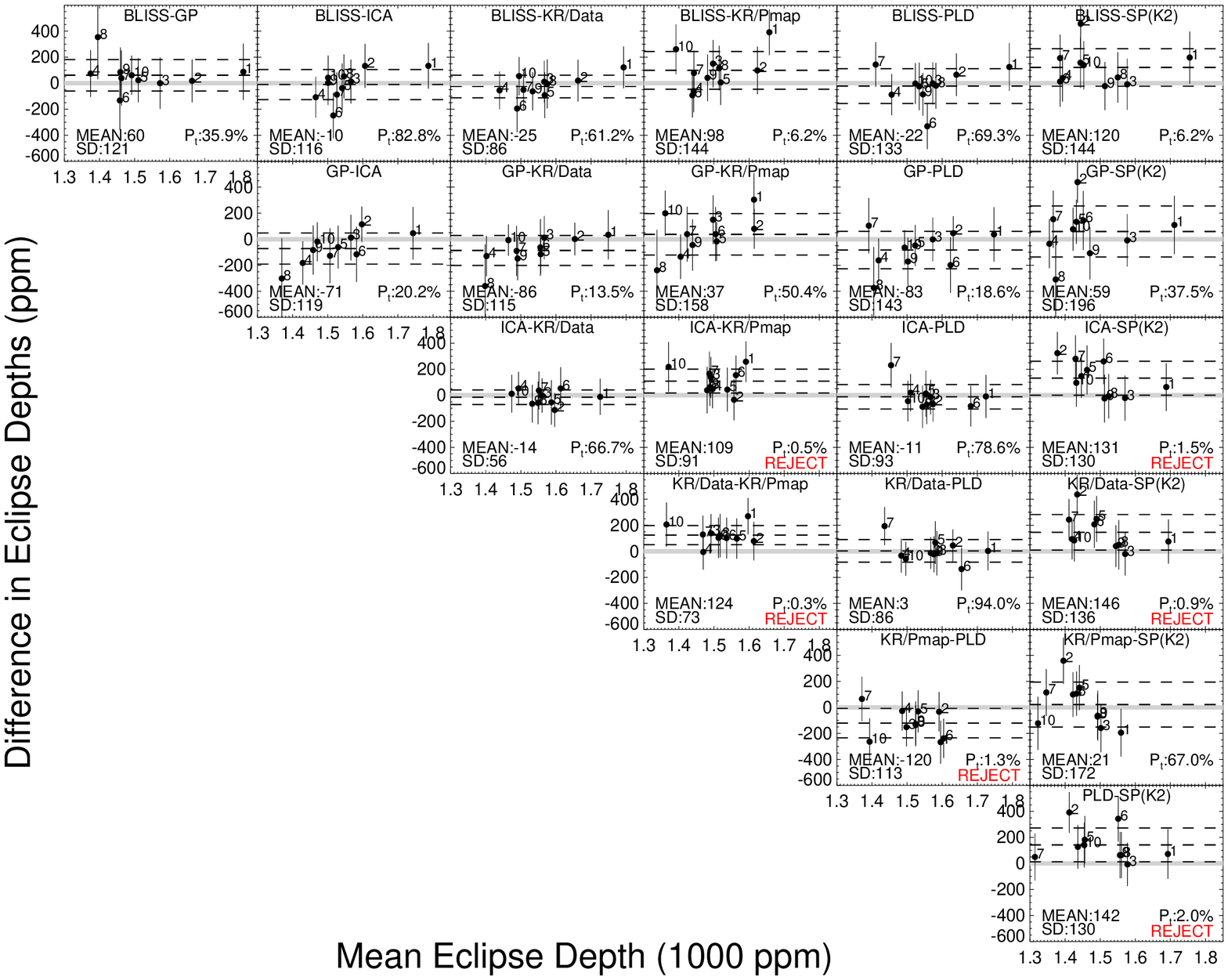}
\caption{Mean/difference plots comparing decorrelation techniques to each other for {\bf real} data.  Each panel plots differences in XO-3b eclipse depths for each epoch for the two techniques given, as a function of the mean depth for the epoch of the pair of techniques.  The epoch is labeled on each point.  Three horizontal dashed lines display the mean difference, or relative bias between the methods, $\pm$ one standard deviation, and the bottom left of each panel prints these numbers.   Horizontal gray lines indicate zero difference.  The bottom right of each panel displays the t-test $p$ value, giving the probability that the $t$ parameter is larger than the measured value if the null hypothesis is true (see text).  If $p<5\%$, the null hypothesis is rejected, which we take to mean that the two techniques are not measuring the same mean eclipse depth.
\label{diff_real_methods}}
\end{figure*}
\begin{figure*}
\plotone{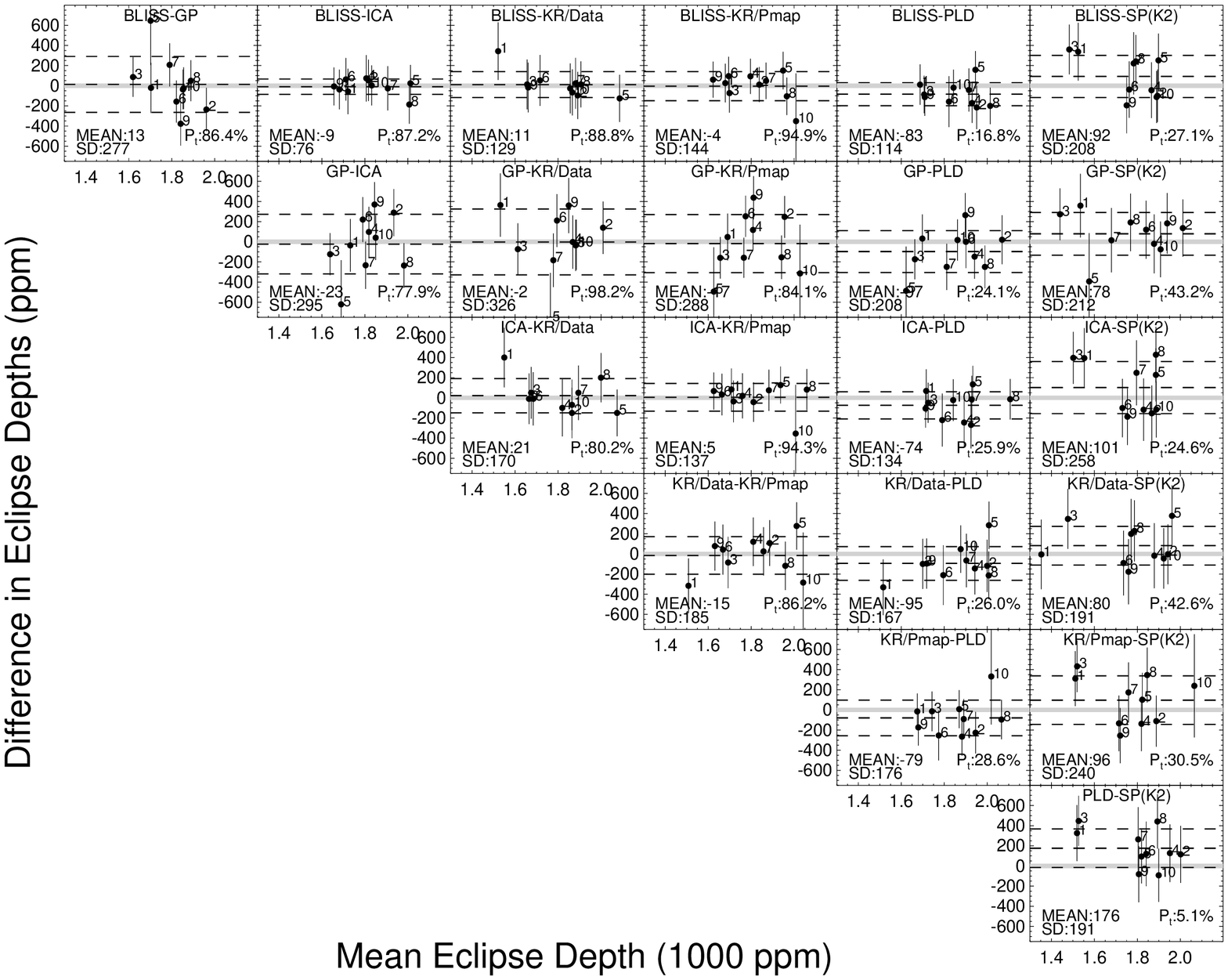}
\caption{Mean/difference plots comparing decorrelation techniques to each other for {\bf simulated} data.  See caption to Figure \ref{diff_real_methods} for more details.
\label{diff_fake_methods}}
\end{figure*}
The repeatability, reliability, and accuracy are all measures applied to the results of a single decorrelation method.  We can also conduct a more direct comparison of methods.  First, we use the mean/difference plotting method of  \citet{Altman:1983bn} to make a visual comparison. Figures \ref{diff_real_methods} (real) and \ref{diff_fake_methods} display these plots for each pair of methods.  Dashed lines show the mean of the differences, $\overbar{\Delta}$, which estimate the relative bias between techniques; and $\pm {\rm SD}(\Delta)$, which bounds the limits of variability.  Column 10 of Tables \ref{eclipse_depth_stats} and \ref{eclipse_depth_stats_sim} list the method that gives the closest match to each of the methods of Column 1.  This was chosen as the method giving the smallest range of eclipse values, min($|\overbar{\Delta}| + {\rm SD(\Delta)}$).

Another way of comparing two approaches is to use the Student's t-test to assess whether the results are drawn from a distribution with the same mean.  The test posits the null hypothesis that both sets of data have the same mean and attempts to reject it.  We use the unpaired version of the test to compute the $t$ statistic (the difference in average values divided by the combined variance) and compare with the t-distribution for the number of degrees of freedom.  The bottom right corner of each panel of Figures \ref{diff_real_methods} and \ref{diff_fake_methods} displays the probability that $t$ is larger than the computed value if the null hypotheses were true.  The null hypothesis is rejected if $p<5\%$, i.e., the measured statistic is in the tail of the distribution.  In all the comparisons for simulated data and most comparisons for real data, the hypothesis is not rejected.  However, for real data, both KR/Pmap and SP(K2) are likely not to have the same mean as ICA, KR/Data, or PLD.

We can also do a global comparison of methods using analysis of variance (ANOVA) F-test, which posits the null hypothesis that {\it all} sets of eclipse depths have the same mean.  This analysis assumes that the group of eclipse depths for each method follows a normal distribution, and that each group has approximately the same variance (usually taken to be within a factor of two of each other, which our measurements satisfy).  Similarly to the t-test, it computes a statistic and compares it with the expected distribution under the null hypothesis.  In this case the statistic is $F$, the ratio of the average variability among groups (the dispersion of group means) to the average variability within groups (average group variance).  The comparison distribution is the F-distribution (also known as the Fisher-Snedecor distribution), which gives the probability of measuring $F$ for the applicable degrees of freedom, given the null hypothesis.  Smaller values of $F$ imply a higher probability that the groups share the same mean. For the real data $F=2.8$, for which only $p=1.6\%$ of the F-distribution has larger values, so we reject the null hypothesis and conclude that not all methods have the same mean.  If we remove KR/Pmap and SP(K2) from the calculation because they were the only methods that had failed t-tests, then $F=0.9$.  In this case 45\% of the distribution has larger values, and we don't reject the null hypothesis.  We conclude therefore that KR/Pmap and SP(K2) eclipse depths are biased relative to the other techniques.  For the simulated data $F=0.8$ (all techniques), for which 57\% of the distribution has larger values, so we don't reject the hypothesis of equal means.

We emphasize that null hypothesis significance tests like {\it t} and {\it F} tests are limited in scope and predictive power.  In particular, they only allow us to {\it reject} the hypotheses of equal means, but not to accept them.  Their probability distributions give the probability, assuming the means are equal, that the corresponding statistic has the measured value, {\it not} the probability that the means are or are not equal given the measured statistic. Nevertheless, they still have value, at least as a first approach to an inter-method comparison.  Bayesian estimation with Monte-Carlo simulations would provide a more robust and comprehensive framework from which to analyze differences and similarities between results \citep[e.g.,][]{Killeen:2005hv,Kruschke:2013jy}, but is beyond the scope of this work.

\section{Discussion}
\subsection{Repeatability and accuracy of IRAC eclipse depth measurements\footnote{This portion leaves out the SP(K2) technique, which was not developed for \Spitzer.}}
We have analyzed 10 real and 10 simulated eclipses of hot Jupiter XO-3b using seven correlated noise removal methods.  The simulations were in some ways an attempt to replicate the real data, but were given larger pointing fluctuations and drifts, thereby increasing correlated noise and decreasing the positional redundancy that many noise removal techniques rely on. 

For the real data, the statistical uncertainties determined on individual eclipse depths accurately describe the scatter in eclipse depths over the 10 visits.  In only one case, BLISS mapping, did the uncertainty need to be increased by 50\%.  For the simulations, all techniques except BLISS mapping required an increase of 20--60\%, implying that the methods may need slight adjustment to allow individual uncertainties to track the increased pointing fluctuations.

We defined three terms relating to measurement stability: repeatability ($R$), the expected difference between repeated measurements; reliability ($r$), the ratio of intrinsic (photon-limited) to measured {\it variability}; and accuracy ($a$), the ratio of intrinsic to measured {\it error}.  Repeatability and reliability are inversely related, and reliability is a normalized estimate of precision.  Accuracy combines both trueness and precision, and can theoretically never have a value less than the reliability.

For real XO-3b data, eclipse depths are repeatable within $R\lesssim 180\,$ppm.  In other words, any two single eclipses are expected to be within 180\,ppm of each other 68\% of the time.  The most repeatable techniques have $R\leq 100\,$ppm, which is about \nicefrac{4}{3} the photon limit ($\sqrt{2}\sigma_{\litl phot}\approx 75$).  For the synthetic data, the repeatability is somewhat larger, $R\lesssim 300\,$ppm.

When comparing the scatter in eclipse depths with the intrinsic uncertainty due to photon noise, all techniques come within a factor of 3 of the photon limit for the real data (reliability $r>0.33$).  ICA and the kernel regression techniques (KR/Data and KR/Pmap) exhibit a scatter consistent with more than two-thirds photon noise ($r \gtrsim 0.65$).  For the simulations, the eclipse depth scatter is within a factor of 2--4 of the photon limit.  Only the BLISS technique had the same value of reliability for the simulations as the real data, whereas the kernel regression techniques showed reductions of 40\%.  Even though BLISS may not be as precise as other techniques in the best circumstances, its precision is the most robust to an increasing positional spread.  

The simulations afforded a unique view into the analysis of eclipses, allowing us to evaluate the accuracy and bias of each method based on knowledge of the true depth.  The root mean-squared eclipse depth error ranged from 2.5 to 5 times the photon noise limit, yielding accuracy values ranging from 0.2 to 0.4.  Most techniques obtained an average eclipse depth within 100\,ppm of the true depth (1875\,ppm).  

We stress that repeatability, reliability, and accuracy are statistics that refer to the quality of  {\it single} measurements.  To say that a technique has a reliability of $r$ means that an individual measurement of the eclipse depth has a 68\% chance of being consistent with other measurements to within $1/r$ times the photon limit (assuming Gaussian statistics).  An accuracy of $a$ means that an individual measurement has a 68\% chance of being within $1/a$ times the photon limit of the true value. Techniques that give lower values of these quantities may nevertheless be extremely accurate when the results are averaged over multiple epochs.  For example, PLD ($r=0.38$ for simulated data) has larger overall scatter in individual measurements than BLISS ($r=0.44$), but because PLD has a much lower bias than BLISS (5 vs -59, when averaged over 10 visits), both techniques have similar values of RMSE and are thus considered equally accurate ($a = 0.39$ for PLD and 0.40 for BLISS).  

\subsection{Is there a ``best'' IRAC correlated noise removal technique?}
After examining the results of processing the $2\times 10$ datasets with seven different techniques for data reduction and eclipse depth measurement, we can make some tentative statements about the relative merits of the methods:
\begin{itemize}
\item When the pointing fluctuations are at a normal level, ICA and the kernel regression techniques (KR/Data, KR/Pmap) return repeatability that is within a factor of $\sim 1.5$ of the photon limit ($r_{\litl real}\ge 0.65$), followed by PLD with $r_{\litl real}\sim 0.5$, BLISS and SP(K2) with $r_{\litl real}\sim 0.4$, and GP with $r_{\litl real}\sim 0.3$. (Here we have used inverse reliability as a normalized proxy for repeatability---see \S3.4.2.) 
\item BLISS is the most precise of all methods when the pointing fluctuations are larger ($r_{\litl sim}\sim 0.4$).
\item The precision of BLISS is the most robust to changes in the pointing fluctuations and drift ($r_{\litl real}=r_{\litl sim}$).  
\item  BLISS, PLD, and ICA are the most accurate and the most reliable (both $a$ and $r \sim 0.4$), at least when pointing fluctuations are larger (simulated data).  
\item PLD (with a quadratic phase curve model) yields the least biased results of all methods (however, all other methods used flat or linear phase curves---see below). 
\item KR/Pmap and SP(K2), both of which did not include phase curve variations in their eclipse fits, return eclipse depths that are strongly biased, for both real data (they are not consistent with having the same mean as the other methods) and simulated data (their measured average biases are more than twice those of most of the other methods).
\end{itemize} 

We emphasize that we have not separately controlled for centroiding, photometry, correlated noise removal, or eclipse depth fitting.  In comparing techniques above, we are really comparing the entire data reduction pipelines that go along with each method.  In particular, the out-of-eclipse phase curve model can significantly bias the measured eclipse depth.  The simulated eclipses have nonlinear time-dependent phase variations that are concave downward (see Fig. \ref{predicted_lightcurve}).  \replaced{One}{Therefore, one} expects eclipse fits using a linear (BLISS, GP, ICA, KR/Data) or flat [KR/Pmap, SP(K2)] phase curve to yield a center-of-occultation flux that is lower than the truth\added{, as indeed seems to be the case}.  

\added{We can calculate the {\it true} eclipse depth bias due to the phase curve model from the (noiseless) input light curve, L(t) (\S\ref{imaging_sec}), by fitting various phase models to the flux outside occultation and measuring the depth.  For the XO-3b simulation shown in Fig. \ref{predicted_lightcurve}, the fit eclipse depth is biased by $-51\,$ppm for a flat phase model, $-27\,$ppm for a linear model, and $-2\,$ppm for a quadratic model.  Not including SP(K2), these values account for approximately 50\% of the measured average biases (Table \ref{eclipse_depth_stats_sim}, column 12).  Given that the uncertainties in the mean depths (Table \ref{eclipse_depth_stats_sim}, column 7) have similar magnitudes to the biases, a larger ensemble of measurements would be necessary to make any definite claims regarding bias.  Nevertheless, much of the {\it true} bias for the methods that used a flat or linear phase model would have been reduced dramatically by a quadratic phase curve.}\deleted{Thus, the eclipse depths would be biased negative (measured depth smaller than true depth), as indeed they are (Table \ref{eclipse_depth_stats_sim}, column 12).}  In the BLISS processing, quadratic and sinusoidal models were tried but were not favored by the Bayesian Information Criterion (BIC; Equation \ref{bicdef}).  The only method whose reported depths are based on a quadratic phase curve, PLD, \replaced{has}{yields} a relatively low positive bias \added{($+5\,$ppm), which is consistent, within its uncertainty, with the expected true bias of $-2\,$ppm}.  

This leads to the question: given that phase variations are expected to be nonlinear (if they exist at all, they are usually periodic), how should we interpret the BIC when it favors linearity?  The BIC often helps minimize free parameters and ensure that models are generalizable among similar datasets; but it also is known to underfit \citep{Dziak:2012um}, not allowing for sufficient variability and sometimes leading to biased results. Another quantitive model selection technique, the Akaike Information Criterion (AIC) tends to {\it overfit} data (allow for too many free parameters) and therefore be too tied to the specifics of a given dataset.  One approach, suggested by \citet{Dziak:2012um} would be to select the best models according to both the BIC and AIC, and bracket a {\it range} of model sizes, instead of specifying definitively one model as the ``best.''  In the end, model selection still requires human judgement to balance quantitative criteria such as the AIC and the BIC with reasonable expectations based on theory.  

\subsection{Are IRAC eclipse depth uncertainties underestimated?}

A recent study by \citet{2014MNRAS.444.3632H} derived systematic uncertainties for IRAC eclipse depths.  They compared 10 2-epoch pairs of \Spitzer\ eclipse depth measurements for six different planetary systems, each epoch measured by different teams, including measurements from three IRAC wavelength bands and one MIPS band, as well as IRAC data taken using both dithers and staring mode.  They estimated the systematic variance in each depth from the squared difference in eclipse depth values between epochs, minus the sum of reported variances (squared uncertainties) for each epoch.  This is equivalent to our estimate of $f_{\litl dis}$ (\S3.2), but for a sample size of $N=2$ instead of 10.  In 5 out of 10 comparisons the difference between epochs was larger than the reported uncertainty by more than a factor of 2.  Combining results across data analysis methods, planetary systems, IRAC wavelength bands, and from both staring and dithering mode, they concluded that in general single eclipse measurements made with {\sl Spitzer}/IRAC either have an uncertainty floor of 500 ppm, or that their uncertainties should be multiplied by a factor of $f_{\litl dis}=3$.  They used their inflated uncertainties to assert that features seen in broadband spectra are more likely due to instrumental systematics than molecular bands. 

Following this, some authors have echoed the conclusions of \citet{2014MNRAS.444.3632H}.  For example, \citet{2015MNRAS.449.4192S} obtained theoretical estimates on the properties of 50 exoplanet atmospheres after first assuming that many of the reported \Spitzer\ eclipse depth uncertainties were underestimated by a factor of 3.  Most recently, a general review on the observation of exoplanet atmospheres \citep{Crossfield:2015jd} also accepted the \citet{2014MNRAS.444.3632H} assertion regarding overestimated \Spitzer\ precision, stating that, ``it is debatable whether broadband photometry usefully determines atmospheric abundances in {\it any} transiting exoplanets [emphasis added].''  If this statement were true, many recent analyses using modern reduction techniques and realistic (but not inflated) uncertainties would be invalidated.  For some examples, see the \citet{2015arXiv151209342W} claims regarding high altitude silicate clouds in WASP-19b and enhanced C/O ratio in HAT-P-7b; or the \citet{2016Natur.529...59S} categorization of the atmospheres of 10 hot Jupiters from clear to cloudy using HST and \Spitzer\ data.

Our conclusions contradict those of \citet{2014MNRAS.444.3632H}.  To avoid the influence of confounding variables that affect measurement stability, the present paper focuses on a single planetary system, using data from a single IRAC band and single observing mode (staring mode), and involves a parallel analysis isolating different correlated noise removal techniques (and their associated data reduction pipelines).  In contrast to the $f_{\litl dis} = 3$ estimate of \citeauthor{2014MNRAS.444.3632H}, we have found for both real and simulated XO-3b data that the statistical uncertainties do not need to be increased by more than 50\% to accomodate the scatter in data [for all decorrelation methods except SP(K2), which was created for K2 and not optimized for \Spitzer], and in many cases no inflation was necessary.  This holds even for simulated data, which had increased correlated noise and decreased spatial redundancy.  Our estimates of $f_{\litl dis}$ include confidence intervals based on 10-epoch samples (column 6 of Tables \ref{eclipse_depth_stats} and \ref{eclipse_depth_stats_sim}), which vary by $\sim\pm\nicefrac{1}{3} f_{\litl dis}$.  As emphasized by \citet{1992JPhA...25.1967L}, the uncertainty on $f_{\litl dis}$ for $N=2$ \citep[the sample size used by][]{2014MNRAS.444.3632H} is much larger, up to a few times the actual value of $f_{\litl dis}$.   

The chief source of the discrepancies between separate eclipse depth measurements examined by \citet{2014MNRAS.444.3632H} is the evolution in both observing and data reduction strategies that has occurred to accomodate exoplanet observation.  One key example of non-repeatability of IRAC eclipse depths cited by \citeauthor{2014MNRAS.444.3632H} is the 4.5\,$\mu$m measurement for HD 209458b.  An early study of this hot Jupiter used broadband {\sl Spitzer} secondary eclipse spectra from 3.6 to 24\,$\mu$m to infer the existence of an atmospheric inversion layer in the planet \citep{2008ApJ...673..526K}.  These 2005 measurements were among the earliest eclipse observations made with IRAC, and were obtained using the (then) standard practice of alternating exposures between each IRAC channel, which required a repointing every $4\times 64$ subarray images.  When \Spitzer\ is commanded to continuously observe an inertially fixed target (``staring'' mode), a source's position will fluctuate over a region of about 0.08\,px diameter in one hour, while also incurring a slow linear drift of about 0.01\,px per hour.  Experience shows that this usually yields sufficient redundancy in source position to decorrelate intra-pixel gain in a set of photometric measurements.  On the other hand, {\sl Spitzer's} blind repointing accuracy is much worse: about 0.3\,px RMS.  It is not surprising, then, that the 2005 measurements of HD 209458b, which were repointed every 256 frames, yielded large discontinuities in the target position, making it extremely difficult to decorrelate the data at 3.6 and 4.5\,$\mu$m and extract accurate eclipse depths (especially using a low order polynomial fit to the intra-pixel gain, as was the common practice).  Subsequent measurements of the full phase curve of HD 209458b (by a team that included two of the three authors on the earlier study) were taken in continuous staring mode with no repointing, and the data were decorrelated using kernel regression as a function of $x$, $y$, and noise pixels \citep{2014ApJ...790...53Z}.  The new methodology resulted in a 35\% lower 4.5\,$\mu$m eclipse depth that did not require an atmospheric temperature inversion.  

\citet[Table 2]{2014MNRAS.444.3632H} use the difference between the 4.5\,$\mu$m eclipse depth derived by \citet{2008ApJ...673..526K} and that derived by \citet{2014ApJ...790...53Z} as a baseline estimate of the {\it systematic uncertainty} in {\sl Spitzer}/IRAC measurements at 4.5\,$\mu$m. This is incorrect, since it treats both approaches to measurement and reduction as equally valid, and equally indicative of the possible range in measurable eclipse depths.  The 2005 IRAC measurements of HD 209458b were taken in such a way as to make the intra-pixel systematics in the InSb arrays virtually uncorrectable. In more recent years observational practice has evolved towards a more optimal staring mode configuration, especially with the 2009 advent of PCRS Peak-Up to ensure that targets are repeatably positioned (to within 0.1\,px) in a region with minimal intra-pixel gain variations \citep{2012SPIE.8442E..1YI}.  Eclipse data taken in this manner eliminate the discontinuous position jumps present in the 2005 data.  

Also, the techniques for removing correlated noise have improved dramatically from the early days of low order polynomial fitting.  Even the sub-optimal 2005 measurements of HD 209458b were shown to be consistent with later measurements after reanalysis using BLISS \citep{2014ApJ...796...66D} and GP \citep{2015MNRAS.451..680E}.  One of the criticisms made by \citeauthor{2014MNRAS.444.3632H} was that reported uncertainties for published eclipse depths were unrealistic and did not sufficiently take systematics into account. We agree that early methods did not adequately estimate the errors, but this is not a problem in most of the newer approaches, as seen in the current paper. 

In his review of the study of exoplanet atmospheres, \citet{Burrows:2014jd} pointed out that observers and theorists have tended to overinterpret the earliest measurements.  The article is a sobering reminder that results from a young field may be overturned by improved approaches to observation, reduction, and theory.  The decrease in \Spitzer/IRAC correlated noise due to staring mode and PCRS Peak-Up, as well as the improved understanding of systematics and development of better decorrelation techniques, have led to a situation in which the variations in eclipse depths described in \citet{2014MNRAS.444.3632H} are now outliers when compared to variations observed today.  \citet{2014MNRAS.444.3632H} is a watershed work that attempted to quantify the uncertainties in \Spitzer\ single exoplanet eclipse depths hinted at by \citet{Burrows:2014jd}, via comparisons between paired studies.  However, like the earliest theoretical conclusions that were biased by outlier eclipse depth {\it measurements}, \citeauthor{2014MNRAS.444.3632H} may have been similarly biased and overinterpreted the earliest {\it variations} in eclipse depths.  

\subsection{Application to future space missions}
Future space missions such as JWST \citep{2008AdSpR..41.1983C} and TESS \citep{2015JATIS...1a4003R}, and proposed missions such as ARIEL \citep{2015DPS....4741620T} and FINESSE \citep{2012SPIE.8442E..41D}, will have similar needs to verify the repeatability and accuracy of their eclipse and phase curve measurements.  These observatories will benefit from having been designed with precision measurements of transiting exoplanets in mind, so the instrumental systematics will not be as significant as for \Spitzer/IRAC, where correlated noise can be as much as 2 orders of magnitude larger than eclipse depths.  However, systematics will still be present in future missions:  JWST will have similar jitter to pixel scale ratios as found in \Spitzer/IRAC \citep{2014PASP..126.1134B}, which will lead to photometric variability due to intra-pixel gain fluctuations.  Furthermore, observers will demand increasingly more precise measurements as more detailed questions are asked regarding e.g., atmospheric variability.  Next generation space observatories will undoubtedly be pushed to the limits of their systematic error budgets and, like \Spitzer, require a thorough assessment of their stability and accuracy.

\section{Conclusions}

We have performed a \Spitzer/IRAC repeatability analysis of 10 real and 10 simulated eclipses of XO-3b using seven correlated noise removal techniques.  Most methods are capable of estimating accurate uncertainties on individual eclipse depths.  The eclipse depth repeatability (expected difference between pairs of measurements) under normal pointing variations averages $\sim 150\,$ppm, only twice the photon limit, but can worsen as the spread in target positions increases.  The BLISS technique, however, is most robust to such changes.  The BLISS, PLD, and ICA techniques are the most accurate and repeatable when the pointing fluctuations are larger.  Future analysis might benefit from separating the phase curve model from the decorrelation technique, as it can bias eclipse depths.

A few recent publications have claimed that \Spitzer\ eclipse depth uncertainties should be increased by a factor of 3.  Such claims rest upon a comparison of literature estimates of varying provenance and quality, using only two epochs per target, and are not substantiated by our more controlled analysis with a larger, more uniform sample.  

Although we have controlled reasonably well for most important observing variables, our conclusions are strictly valid only for the IRAC 4.5\mum\ array, and in the particular signal-to-noise regime of XO-3b (photon noise limit on an eclipse depth of $\sim 50\,$ppm).  As multi-epoch {\sl Spitzer}/IRAC measurements accumulate for a variety of exoplanet targets, the data will better support more broad-based repeatability analysis, which will constrain further the limits of variability for reduction techniques, and ultimately for the instrument itself.  

Some of the lessons learned with IRAC can be usefully applied to future space missions.  The high degree of repeatability demonstrated in this paper was facilitated by a careful characterization and optimization of pointing during exoplanet observations \citep{Grillmair:2012fq,2012SPIE.8442E..1YI}.  This understanding of the systematics was greatly facilitated by a set of dedicated calibration observations.  The IRAC team has also found that hosting exoplanet data workshops and engaging the active research community has led to the optimization of observing strategies and improved the quality of data greatly. This paper shows that state of the art reduction techniques do an excellent and consistent job of mitigating systematic noise.  Focused data challenges could prove equally effective for future exoplanet space missions.

\acknowledgments

We thank the anonymous referee for comments that greatly improved the quality of the paper.  We thank Nick Cowan for a thorough reading of an early version of the manuscript and for discussions that helped clarify the conclusions. This work is based on observations made with the \Spitzer\ Space Telescope, which is operated by the Jet Propulsion Laboratory, California Institute of Technology under a contract with NASA.  G. Morello received support from ERC project number 617119 (ExoLights).  K. Stevenson recognizes support from the Sagan Fellowship Program, supported by NASA and administered by the NASA Exoplanet Science Institute (NExScI).  This research has made use of the Exoplanet Orbit Database and the Exoplanet Data Explorer at \url{exoplanets.org}.



\facility{Spitzer(IRAC)}

\software{IRACSIM (\url{http://dx.doi.org/10.5281/zenodo.46270}), IDL}



\appendix

\section{IRACSIM: An IRAC Data Simulator for Point Source Images}
To produce the simulated XO-3b observations used for the Data Challenge, we used {\tt IRACSIM}\footnote{\url{http://dx.doi.org/10.5281/zenodo.46270}}, a package built in the IDL programming language.  The program uses a model of the \Spitzer/IRAC system to create synthetic IRAC point source measurements, outputting FITS image (or image cube) files similar to those produced by the IRAC basic calibrated data (BCD) pipeline. The simulator model is built on three major components of \Spitzer/IRAC behavior: (1) pointing, (2) imaging, and (3) Fowler sampling.  We give an overview of this model here.

\subsection{Pointing}
The IRAC pointing model specifies the position of a point source as a function of time, ($x[t],y[t]$).  The model has four main components, based on the known structure of \Spitzer\ pointing variations \citep{Grillmair:2012fq}: a high frequency fluctuation or ``jitter'' with amplitude $\sim 0.05\,$px; a sawtooth-shaped ``wobble'' due to a battery heater cycling on and off (period $\sim 40$\,min, amplitude $\sim 0.05\,$px); an approximately 30 minute initial drift of up to 0.1\,px; and a long term drift of $\sim 0.3\,$px per day.  \citep[See also][Figure 8 for high fidelity measurements of jitter, wobble, and drift.]{2014ApJ...793..120H}
The pointing as a function of time is given by:
\begin{eqnarray}
x(t) & = & x_j(t) + x_w(t) + x_{\litl sd}(t) + x_{\litl ld}(t); \\
y(t) & = & y_j(t) + y_w(t) + y_{\litl sd}(t) + y_{\litl ld}(t). 
\end{eqnarray} 

The jitter component is the sum of a sine wave plus a randomly generated $1/f$ noise:
\begin{eqnarray}
x_j(t) & = & A_j\sin[2\pi(t-t_0)/P_j + \phi_j]\,\cos(\theta_j) + {\bf FBM}(A_{\litl fbm},\beta,t)\\
y_j(t) & = & A_j\sin[2\pi(t-t_0)/P_j + \phi_j]\,\sin(\theta_j) + {\bf FBM}(A_{\litl fbm},\beta,t)
\end{eqnarray}
Here, $A_j$ is the jitter amplitude; $t_0$ is the time of the last spacecraft pointing reset, usually via PCRS Peak-Up; $P_j$ is the jitter period; $\phi_j$ is the phase shift of the jitter; and $\theta_j$ is the ``axis'' of the jitter, the angle on the pixel grid (with respect to the $x$ axis) over which the sinusoidal component of jitter oscillates.  The term ${\bf FBM}(A_{\litl fbm},\beta,t)$ is a random variable representing a fractional brownian motion noise with power spectral index proportional to $f^{1/\beta}$ and having peak amplitude $A_{\litl fbm}$, constructed according to the prescription of \citet[][Section 4]{Stutzki:1998vy}.

The wobble component is modeled as a ``skewed sinusoid'':
\begin{equation}
w(t) = A_w(t)\,\sin[2\pi(t-t_0)/P_w(t) + \phi_w + \phi_{\litl sk}(t)],
\end{equation}
where $A_w(t)$ is the amplitude of the wobble, $P_w(t)$ is the period, $\phi_w$ is a constant phase shift, relative to $t_0$, and $\phi_{\litl sk}(t)$ is an additional phase shift that varies with time, giving $w$ its skewed shape. Let $q(t) \equiv (t-t_0)/P_w(t) + \phi_w/(2\pi) \pmod{1}$.  The skew phase function is:
\begin{equation}
\phi_{\litl sk}(t) = \left\{
\begin{array}{ll}
\pi\left(\frac{1}{2S_w}-2\right)q &: 0\le q < S_w\\
\pi\left(\frac{q-S_w}{1-2S_w}-2q + \frac{1}{2}\right) & :S_w \le q < 1-S_w\\
\pi\left(\frac{1}{2S_w}-2\right)(q-1) & : 1-S_w\le q < 1.
\end{array}
\right.
\end{equation}

Here, $S_w$ is the phase of the peak amplitude (as a fraction of the period), which defines the amount of skewness.   In a normal sine wave, $S_w$ equals 1/4, i.e., the curve peaks when the argument of the sine equals $\pi/2$. If $0<S_w<1/4$ the curve has a faster than sinusoidal rise, and is skewed to the left.  If $1/4<S_w<1/2$ the curve has a slower than sinusoidal rise and is skewed to the right.  Either set of $S_w$ choices results in a smoothly varying sawtooth-like curve.  Additional flexibility is enabled by a time variable wobble amplitude $A_w(t)$ and period $P_w(t)$, with the values varying continuously in a random walk having maximum excursions assignable via the parameters $\Delta A_{w,{\litl max}}$ and $\Delta P_{w,{\litl max}}$.  One final parameter that specifies the $x$ and $y$ projections of the wobble is the axis, $\theta_w$:
\begin{eqnarray}
x_w(t) &=& w(t)\cos(\theta_w)\\
y_w(t) &=& w(t)\sin(\theta_w).
\end{eqnarray}

The short term drift appears to have periodic and asymptotic behavior, so we model it with a rapidly decaying sinusoid:
\begin{equation}
s(t) = \frac{A_{\litl sd}}{\sin(\phi_{\litl sd})} \sin[2\pi(t-t_0)/P_{\litl sd} + \phi_{\litl sd}]\,\exp[-(t-t_0)/\tau_{\litl sd}],
\end{equation}
where $A_{\litl sd}$ is the ``asymptotic decay,'' the difference between the initial ($t=t_0$) and final ($t\rightarrow\infty$) values of the function; $\phi_{\litl sd}$ is the phase of the sinusoid; $P_{\litl sd}$ is its period; and $\tau_{\litl sd}$ is the decay time.  The short term drift is projected along the axis, $\theta_{\litl sd}$, onto the pixel grid:
\begin{eqnarray}
x_{\litl sd}(t) &=& s(t)\cos(\theta_{\litl sd})\\
y_{\litl sd}(t) &=& s(t)\sin(\theta_{\litl sd}).
\end{eqnarray}

Finally, the long term drift is a simple linear function of time:
\begin{eqnarray}
x_{\litl ld}(t) &=& A_{\litl ld}\,(t-t_0)\,\cos(\theta_{\litl ld})\\
y_{\litl ld}(t) &=& A_{\litl ld}\,(t-t_0)\,\sin(\theta_{\litl ld}),
\end{eqnarray}
where $A_{\litl ld}$ is the drift rate and $\theta_{\litl ld}$ is the axis of projection.

Table \ref{ptg_params} lists the range of inputs to the pointing model used in simulating the XO-3b data. We chose not to duplicate exactly the pointing fluctuations as observed in the real dataset, but attempted to simulate a range of possible \Spitzer\ observing conditions, and thus a range of possible decorrelation situations.  To do this, most of the parameters for a given epoch were generated randomly within the predefined ranges given.   

\begin{deluxetable}{rllc}
\tabletypesize{\scriptsize}
\tablecaption{Pointing Model Parameter Ranges\label{ptg_params}}
\tablewidth{0in}
\tablehead{
\multicolumn{2}{r}{Parameter} & \colhead{Range\tablenotemark{a}} & \colhead{Type} 
}
\startdata
Jitter & $A_j$ & 0.04\,px & C\\
       & $P_j$    & 60\,s & C \\
       & $\phi_j$     & 0 to 2$\pi$\,rad & U\\
       & $\theta_j$      & -45$\arcdeg$ & C \\
       & $A_{\litl fbm}$ & 0.4\,px & C \\
       & $\beta$         & 1 & C \\
Wobble & $A_w$ & 0.018 to 0.034\,px & U\\
       & $P_w$    & 1200 to 2800\,s & U\\
       & $\phi_w$     & -1 to 1\,rad & U\\
       & $S_w$      & 0.1 to 0.4 & G\\
       & $\Delta A_{w,{\litl max}}$ & 0.01\,px & C\\
       & $\Delta P_{w,{\litl max}}$ & 10\,s & C \\
       & $\theta_w$      & -80 to -45$\arcdeg$ & G\\   
Short Term Drift & $A_{\litl sd}$ & 0 to 1\,px & U \\
                 & $P_{\litl sd}$ & 395.6\,s & C \\
                 & $\phi_{\litl sd}$ &  $7\pi/4\,$rad & C \\
                 & $\tau_{\litl sd}$ & -1800 to 1800\,s & U\\
                 & $\theta_{\litl sd}$  & 100$\arcdeg$ & C\\
Long Term Drift  & $A_{\litl ld}$ & 0 to 0.0208$\arcsec\,\rm{hr}^{-1}$ & U\\
                 & $\theta_{\litl ld}$ & -95 to -55$\arcdeg$ & U\\
\enddata
\tablenotetext{a}{Ranges give either hard limits of a uniform deviate (``U'' in column 3), $\pm 1\,\sigma$ of a Gaussian deviate (``G'' in column 3), or a constant value (``C''in column 3).}
\end{deluxetable}

\subsection{Imaging\label{imaging_sec}}
After using the IRAC pointing model to predict the position of a point source as a function of time, the IRAC Point Response Function (PRF) allows one to compute the image of the source at each of those positions.\footnote{The core PRFs from the cryogenic mission are packaged along with IRACSIM.  They can be downloaded separately at \url{http://irsa.ipac.caltech.edu/data/SPITZER/docs/irac/calibrationfiles/psfprf/}.  See also the IRAC Instrument Handbook, section C.1.}  The PRF is essentially a convolution of the optical point spread function (PSF) and the intra-pixel response function, sampled on each of the IRAC detector arrays. There are 25 PRF image files per IRAC array, each computed for a different region of the array.  The files in turn contain $5\times 5$ interleaved sets of point source realizations offset 1/5 pixel from each other.  For a point source at a given ($x[t],y[t]$) decimal pixel location, the image of the source at pixel $i_pj_p$ is made by interpolating between the $5\times 5$ PRF realizations
\begin{equation}
I(i_p,j_p,t) = {\rm PRF}_{i_pj_p}^{\litl interp}(x[t],y[t]\/).
\end{equation}

Since the core PRF files currently available were built from cryogenic data, we have converted them to post-cryogenic IRAC by assuming that the structure of a point source image is the same as in the cryogenic mission (the optical PSF is unchanged), but that the intrapixel response has changed.  To account for this, we scaled the 25 cryogenic PRF realizations {\it with respect to each other} such that aperture photometry varied according to the measured post-cryogenic photometric gain map at the same intrapixel offsets.  In addition, all PRF centers were shifted such that the center of light centroid (Equations \ref{x_centroid} and \ref{y_centroid}) yields the correct result at zero pixel phase.  

The absolute scaling of $I(i_p,j_p,t)$ is arbitrary.  We rescale it to electron flux using (1) an input desired aperture flux for the point source, $f_{\litl ap}(r_{\litl ap})$ (Jy), (2) an aperture radius $r_{\litl ap}$ (px) for which the flux will be obtained, (3) a normalized light curve specifying the relative flux variations, $L(t)$ and (4) a scaling relationship giving the number of photoelectrons per second in the peak image pixel, divided by the flux in a 3-pixel aperture, $E_{\litl peak}(3) \equiv \dot{e}_{\litl peak}/f_{\litl ap}(3)$. If ${\rm PRF}^{\litl peak}$ is the peak value in the set of PRF images and $a(r)$ is the aperture correction in an aperture of radius $r$, the rate of photoelectron production in each pixel is 
\begin{equation}
\dot{e}(i_p,j_p,t) = L(t)\,I(i_p,j_p,t)\,\frac{E_{\litl peak}(3)}{{\rm PRF}^{\litl peak}} \frac{a(r_{\litl ap})}{a(3)}\,f_{\litl ap}(r_{\litl ap}).
\end{equation}

\subsection{Fowler Sampling\label{fowler_section}}
\begin{figure}
\plotone{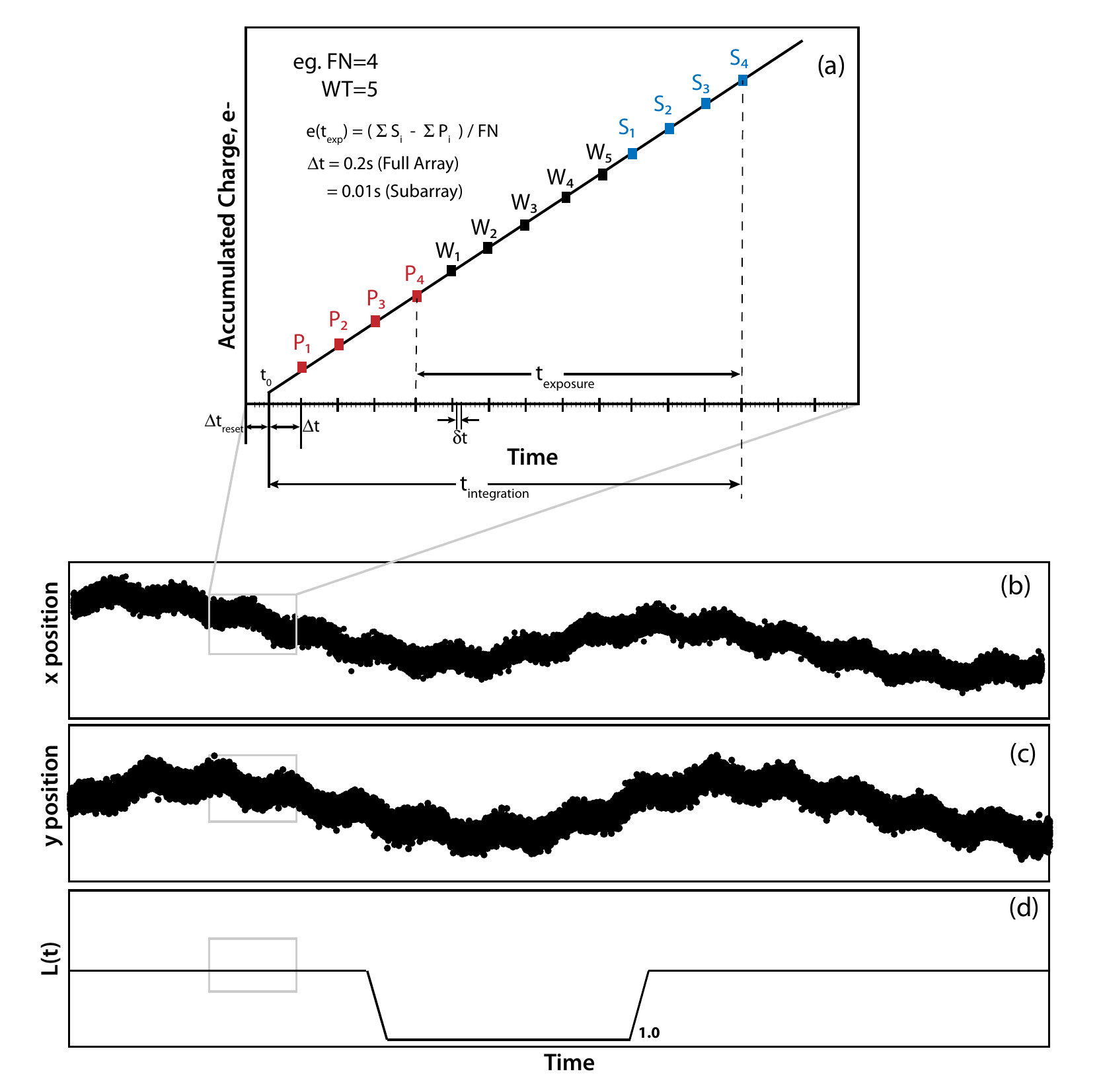}
\caption{Schematic diagram showing the pointing and sampling aspects of the IRAC simulator. Panel ({\it a}) shows the charge on a pixel as a function of time during an IRAC measurement.  We indicate the location of the Fowler sampling ``pedestal'' $P_i$ and ``sample'' $S_i$ measurements, for $FN=4$ and $WT=5$.  (This sampling yields a nonstandard frame time of 2.6\,s, shown here only for illustrative purposes.)  The IRAC sample time is shown as $\Delta t$, and the time resolution of the simulation $\delta t$ is indicated. Panels ({\it b}) and ({\it c}) indicate the time evolution of the $x$ and $y$ pixel position of a point source on an IRAC array.  Panel ({\it d}) displays the raw lightcurve for an unresolved eclipsing planetary system $L(t)$. Gray boxes on panels ({\it b}), ({\it c}), and ({\it d}) show where the pixel sampling in panel ({\it a}) takes place.\label{IRAC_lightcurve_model}}
\end{figure}

Given $\dot{e}(i_p,j_p,t)$, a function that can be evaluated at arbitrary time, we produce a simulated IRAC image by mimicking the integration and sampling properties of the IRAC electronics.  

IRAC acquires data using the Fowler-sampling technique, defined by the sample time, $\Delta t$, the Fowler number, FN, and the Wait Ticks, WT (IRAC Instrument Handbook, section 2.4).  The sample time $\Delta t$ is fixed at 0.2\,s for full array readout and 0.01\,s for subarray readout.   At the beginning of an IRAC measurement, each detector (pixel) is reset.  Charge is then accumulated due to photoelectron production and noise.  The accumulated charge in a pixel is read out every $\Delta t$ seconds, for FN ``pedestal'' reads, $P_i(t)$, WT ``wait'' samples are skipped, and FN ``signal'' reads, $S_i(t)$ are measured.  Figure \ref{IRAC_lightcurve_model} ({\it a}) is a schematic depiction of Fowler sampling, and its relationship with the pointing model. 

Each data file contains either one $256\times 256$-pixel image for full array readout mode, or 64 $32\times 32$-pixel images for subarray readout.  Define $\delta t$ as the rate at which the pointing model is sampled.  The total charge accumulated in one pointing sample at time t is $e(i_p,j_p,t) = \dot{e}(i_p,j_p,t)\delta t.$  To capture possible rapid fluctuations in pointing that might affect the stellar image over the integration, we let the Fowler sample time $\Delta t$ be somewhat larger than $\delta t$.   We typically use $\delta t = \Delta t/10$, so there are $n=10$ PRF realizations to be integrated per Fowler sample.  

We compute $e(i_p,j_p,t)$ every $\delta t$ seconds over the course of the entire integration, which lasts $t_{\litl int} = [2({\rm FN}) + {\rm WT}]\Delta t = [2({\rm FN}) + {\rm WT}]n\delta t$ seconds.  The number of total model samples in the integration is therefore $N{\litl samp} = [2({\rm FN}) + {\rm WT}]n$.  

The accumulated charge is stored for {\it every} Fowler sampling interval (starting at $P_1$ and ending at $S_{\litl FN}$), including the wait ticks for proper noise accumulation.  For the $k$th Fowler sampling interval, the total accumulated charge is 
\begin{equation}
e_k^{\litl mean}(i_p,j_p) = e_{k-1}(i_p,j_p) + \sum_{l=1}^{n} e(i_p,j_p,t_{kl}),
\end{equation}
where $e_0(i_p,j_p)\equiv 0$ and $t_{kl}$ is the time of the $l$th pointing model subsample of the $k$th Fowler sampling interval.  The superscript ``mean'' indicates that this is an estimate of the mean charge.  The actual electron counts will vary due to counting noise, and this is modeled via a Poisson random deviate {\bf P}:
\begin{equation}
e_k(i_p,j_p) = {\mathbf P}[e_k^{\litl mean}(i_p,j_p)].
\end{equation}
Here {\bf P}$[\mu]$ indicates a Poisson random variable with mean $\mu$.  

We separate out Fowler pedestal and signal samples by realizing that the $i$th pedestal read is overall sample $i$, whereas the $i$th signal read is overall sample (FN + WT + $i$).  We also note that each time we read the detectors, we must add readout noise, which we model as a Gaussian random variable {\bf G}$(\mu=0,\sigma=\sigma_{\litl RN})$.  The readout noise standard deviation, $\sigma_{\litl RN}$, is listed in Table 2.3 of the IRAC Instrument Handbook.\footnote{\url{http://irsa.ipac.caltech.edu/data/SPITZER/docs/irac/iracinstrumenthandbook/7/}}  Therefore,
\begin{eqnarray}
P_i(i_p,j_p) & = & e_i(i_p,j_p) + {\mathbf G}(0,\sigma_{\litl RN});\\
S_i(i_p,j_p) & = & e_{({\litl FN} + {\litl WT} + i)}(i_p,j_p) + {\mathbf G}(0,\sigma_{\litl RN}).
\end{eqnarray}

The result of Fowler sampling is an image which measures the mean electron counts accumulated over the ``exposure time,'' $t_{\litl exp} \equiv ({\rm FN}+{\rm WT})\Delta t$, or the time between the $i$th pedestal and the $i$th signal:
\begin{equation}
e(i_p,j_p) = \sum_{i=1}^{\litl FN} \frac{S_i(i_p,j_p) - P_i(i_p,j_p)}{\rm FN}.
\end{equation}
The analog-to-digital converter of the IRAC electronics measures photoelectron accumulation in terms of digital data number (DN) via the proportionality DN$=e$/GAIN, where GAIN$\approx 3.7$.  Then, the SSC data pipeline produces basic calibrated data (BCD) images in units of \mjysr :
\begin{equation}
{\rm BCD}(i_p,j_p) = \frac{({\rm FLUXCONV})\times e(i_p,j_p)}{{\rm GAIN}\times t_{\litl exp}}.
\end{equation}
Here FLUXCONV is the flux conversion factor between \mjysr\ and ${\rm DN}\,{\rm s}^{-1}$ derived by the SSC.\footnote{See \url{http://irsa.ipac.caltech.edu/data/SPITZER/docs/irac/warmimgcharacteristics/} for the values of FLUXCONV for the InSb arrays.}  The {\tt IRACSIM} package produces images (and image cubes, for subarray measurements) in BCD units.

\subsection{Input and Output}
In addition to the pointing model parameters, {\tt IRACSIM} accepts the following inputs: (1) the position(s) of one or more point sources (in either celestial or pixel coordinates; if positions are given in celestial coordinates then a reference coordinate and its pixel position must also be given); (2) date and time of observation; (3) the source flux density in an aperture, $f_{\litl ap}$ (and the aperture optionally); (4) a source light curve $L(t)$; and (5) the full set of observational parameters allowable in the \Spitzer\ Planning Observations Tool (Spot).  (For example: the instrument channel number, frame time, number of repeats, full or subarray readout).

The output of the program is a facsimile of the output of a real \Spitzer/IRAC observation: a set of BCD image files and uncertainty files, with realistic FITS headers containing standard time and astrometry information that is correct for the simulated observation.  We also add history items, comments, and new keywords that are specific to the simulation.  For example, the mean pixel location of the target throughout the integration is printed in the header.

\subsection{Exoplanet wrapper}
 An additional wrapper subroutine has been written to accomodate simulation of exoplanet measurements with {\tt IRACSIM}.  The wrapper features realtime access to the \url{Exoplanets.org} database of planetary system parameters \citep{2014PASP..126..827H}. Its main job is to create model exoplanet phase curves as input light curves $L(t)$ to the IRAC simulator.  It uses the thermal phase variations model of \citet{2011ApJ...726...82C}, and the transit (and eclipse) shape model of \citet{2002ApJ...580L.171M}, allowing for the effects of nonlinear limb darkening in the transit.  The specifics of \Spitzer\ recommended exoplanet observational practice are built in: long AORs are broken into 12\,hr pieces, with a 30-minute settling AOR at the beginning, and the enhanced accuracy of target centering with PCRS Peak-Up is simulated.
 
 \section{Description of Correlated Noise Removal Techniques}
We review below the seven techniques for removing correlated noise used to reduce the XO-3b datasets described in this paper, adding specific notes on implementation.

\subsection{BiLinearly-Interpolated Subpixel Sensitivity (BLISS) mapping}
BLISS mapping \citep{2012ApJ...754..136S} uses bilinear interpolation over a photometric dataset, to predict the intra-pixel response at a given ($x,y$) location.  The procedure establishes a subpixel rectangular grid of node points, referred to as ``knots,'' spanning the dataset.  Each knot is assigned the mean flux value from among all points in the dataset for which that knot point is the nearest.  The intra-pixel gain at a given data point is then computed from the knot fluxes via bilinear interpolation to the point ($x,y$).  

For the implementation described here, performed by H. Diamond-Lowe and K. Stevenson, photometric measurements were obtained using the POET pipeline described in \citet{2012ApJ...754..136S}, which produced artifact-corrected BCD images interpolated to a 1/5-pixel grid. Centroid positions were measured by fitting a 2D Gaussian profile with fixed width \citep[see the Supplemental Information for][]{2010Natur.464.1161S} on the resampled images, and fluxes were measured using aperture photometry.  Intra-pixel effects were removed using BLISS mapping, and various models were attempted to fit the decorrelated light curve, including a flat or possible linear detector ``ramp'' (time-dependent flux baseline) and flat, linear, quadratic, or sinusoidal phase variations.  The eclipse depth, duration, and time of ingress and egress were fit separately for each epoch, as well as commonly among all visits.  Acceptance of model parameters was decided by minimizing the Bayesian Information Criterion (BIC):
\begin{equation}
{\rm BIC} = -2\,\ln \hat{L} + k\,\ln(n),
\label{bicdef}
\end{equation}  
where $\hat{L}\equiv p(x|\hat{\theta},M)$ is the maximized value of the likelihood function of the data $x$ given the maximizing parameters $\hat{\theta}$ and the model $M$, $k$ is the number of free parameters, and $n$ is the number of data points in $x$.  A Differential Evolution Markov Chain (DE-MC) routine \citep{terBraak:2008iw} was used to explore the phase space of parameters and estimate their uncertainties \citep[for details, see][]{2012ApJ...754..136S}.  

\subsection{Gaussian process regression (GP)}
Gaussian process regression is a procedure for using the correlation properties of a dataset to predict the value at an arbitrary point.  It is alternately known as Kriging and Wiener-Kolmogorov prediction, and was first described in the astrophysical literature by \citet{1992ApJ...398..169R} as a means of interpolating irregularly spaced data.  The technique was used to model instrumental systematics in exoplanet observations by \citet{2012MNRAS.419.2683G} and first applied to \Spitzer/IRAC data by \citet{2015MNRAS.451..680E}.   

For the Data Challenge, the GP analysis by T. Evans started with a maximum likelihood fit to the eclipse depth, mid-eclipse time, and the variance in the white noise, plus a set of parameters for kernel functions describing how the {\it covariance} between two photometric measurements varies with their distance in pixel ($x,y$) and time.  The covariance kernel functions are used analogously to the kernel regression function described below, except the standard kernel regression is applied directly to the photometry.  Uncertainties for the eclipse parameters were obtained using Markov Chain Monte Carlo (MCMC) with Metropolis-Hastings sampling in the region of maximum likelihood.  In the final MCMC step the covariance kernel parameters and white noise variance were held fixed to allow rapid evaluation of the likelihood.  One drawback to the GP method is that the evaluation of the $N\times N$ empirical covariance matrix among $N$ data points is often prohibitive with large datasets.  To avoid this difficulty, fluxes and centroids were binned as a function of time in groups of $\sim 30\,$s, resulting in $N\sim 1000$ data points in each eclipse light curve \citep[see][for more details]{2015MNRAS.451..680E}.

\subsection{Independent component analysis (ICA)}
Independent component analysis is a non-parametric technique for separating blended signals, with little specific {\it a priori} knowledge of their structure.  This is the classic ``cocktail party problem'' of signal processing, which attempts to mimic the human brain's innate capacity for hearing multiple speakers in a crowded room \citep{Hyvarinen:2000ho}.  In contrast to {\it principle} component analysis, ICA does not assume that the statistically independent signals follow Gaussian distributions, and in fact attempts to maximize the {\it non}-Gaussianity after separation.  The methods of ICA were first developed for exoplanet light curve analysis by \citet{2012ApJ...747...12W} and used on \Spitzer\ data by \citet{2014ApJ...786...22M}.

The ICA data reduction of the XO-3b real and simulated eclipse datasets, by G. Morello, used a new ``wavelet-pixel'' variant on the approach introduced in \citet{2014ApJ...786...22M} and \citet{2015ApJ...808...56M} for transits.  In this variant, the source separation operates on wavelet-transformed individual pixel light curves, after which the resulting components are transformed back to the time domain.  The wavelet transform was useful for enhancing the signal to noise ratio in the lower frequency instrument systematics components prior to ICA.   By operating on the individual pixel light curves, ICA circumvents a built-in degeneracy that occurs for most decorrelation techniques, which decorrelate aperture fluxes using ($x,y$) centroids.\footnote{Since flux and centroid are both weighted sums of pixel intensities (center-of-light centroids are linear sums, whereas Gaussian fits are effectively nonlinear sums), flux and centroid are always correlated {\it by definition}.  This intrinsic correlation effectively adds ``noise'' to the flux vs. centroid signal. }  

The sum of an eclipse light curve model (including a linear phase variation) and scaled versions of the non-eclipse independent components was then fit to the raw light curve.  An Adaptive Metropolis MCMC algorithm with delayed rejection produced chains of 300,000 values to serve as samples of the posterior distributions of the fit parameters.  These distributions yielded estimates of the parameters and their uncertainties.  The final error bars were then increased to include the ICA component separation error.  A full description of the implementation of ICA on the XO-3b real dataset is given by \citet{Morello:2016wq}.

\subsection{Kernel regression (KR/Data, KR/PMap)\label{KRsection}}
Kernel regression is the first nonparametric technique used to measure and correct the intra-pixel sensitivity of the \Spitzer/IRAC InSb detectors.  In mathematics and engineering, the general use of kernel-based methods was originally applied to the estimation of density functions (e.g., histograms).  Eventually they were proposed as potential tools for regression (i.e., the fitting or prediction of function values) \citep{Nadaraya:2006de,Watson:1964kb}.  The kernel regression estimator is a weighted average of the measured data, with a kernel function specifying how the weight decreases with distance from the target point ${\mathbf x}=(x,y,...)$ to be estimated.  Limiting the contributing data points to the $k$ nearest neighbors to the target is an additional expedient for faster computation \citep{Stone:1977ey}.  The first application of kernel regression to estimate the intra-pixel response in \Spitzer\ photometry was done by \citet{2010PASP..122.1341B}.  The use of the Noise Pixel parameter, $\tilde{\beta}$, as a third component of the distance metric of the weighting kernel (in addition to $x$ and $y$ pixel centroid) was first described by \citet{2013ApJ...766...95L}.

The most commonly used version of kernel regression, KR/Data, uses the data to be corrected as its own ``training set,'' i.e., the data (except the single datum being corrected) are used in the kernel average to obtain the correction.  This requires that the observations contain sufficient redundancy in positioning to allow estimation of its own correlated noise via the inverse distance-weighted average, even in the presence of temporal variations in the astrophysical source.  The published reduction of the real XO-3b eclipse dataset, described in \citet{2014ApJ...794..134W}, used KR/Data.  A complementary analysis of the synthetic XO-3b dataset was performed by I. Wong for the Data Challenge.  For both analyses only the $x$ and $y$ centroids (as measured using the center of light technique; see \S3.1) were used in the kernel's distance metric, but $\tilde{\beta}$ was employed for most eclipses as a scale factor in determining the optimal aperture size for the photometry. \citeauthor{2014ApJ...794..134W} chose $k=50$ nearest neighbors for the weighted sums.  They fit the data in two ways: each epoch separately, and all epochs combined.  The separate fits were only concerned with the eclipse depth, time of mid-eclipse, and linear slope of phase curve, whereas the global fits also included the planet-to-star radius ratio, the orbital inclination, and the semi-major axis to stellar radius.  Both fits were performed using a Levenberg--Marquardt (L-M) nonlinear least squares algorithm. They then used both a prayer-bead method and an MCMC routine to estimate the distributions of each parameter and their uncertainties, and reported the largest uncertainty of the two methods. 

A variation on the kernel regression technique, KR/Pmap, uses the photometry of a separate calibration star as the training set for the regression (Krick et al., in press).  For each science data point to be corrected, the $k$ nearest neighbors in the pixel mapping (pmap) dataset are found, based on the Euclidean distance in $x$ and $y$ centroid and $\tilde{\beta}$.  Similarly to the KR/Data implementation, $k=50$ was chosen.  The kernel-weighted pmap data are then summed and normalized by the calibration star flux averaged over the pixel.  The potential benefit of KR/Pmap over KR/Data is that the correction is not built from the science measurements themselves, and therefore time-varying astrophysical signal does not contribute to the kernel averages.  On the other hand, detector variability (e.g., latent charge) may differ between the calibration star measurements and those of the data to be corrected, and bias the regression.

The KR/Pmap analysis of the Data Challenge measuremements was performed by J. Krick and J. Ingalls.  Different calibration datasets were used to correct the real and synthetic XO-3b measurements.  For the real data, the {\sl Spitzer} Science Center (SSC) has accumulated approximately 400,000 pixel mapping measurements of BD+67 1044, a star that is not known to vary, which are positioned near the ``sweet spot'' (peak of response) of the Ch 2 (4.5\mum) subarray pixel (15,15). Since the IRAC simulator uses an idealized PRF that cannot replicate the detailed structure of the actual pixel response, the real BD +67 1044 dataset is not appropriate for reduction of simulated data.  Instead, a synthetic pixel mapping set was created.  The measurements were designed to mimic the actual pmap measurements of BD +67 1044, with similar source flux, integration parameters, mapping centers, and number of data points.  The pointing model parameters were taken from the same ranges as for the XO-3b simulations (Table \ref{ptg_params}), to approximate realistic motions during integration and sampling. 

Eclipse parameters were derived from the KR/Pmap-decorrelated data by fitting a \citet{2002ApJ...580L.171M} light curve shape with no phase trend using an L-M nonlinear least squares algorithm.  The uncertainties returned were solely the formal uncertainties in the L-M fit, and should be considered underestimates.  As a check on the results, Transit Analysis Package \citep[TAP;][]{2012AdAst2012E..30G} was also used to fit the eclipses, after setting the limb darkening coefficients to zero.  While the uncertainties tended to be more realistic under TAP's MCMC analysis, the eclipse depths themselves were systematically low compared to both the L-M fit and the mean of the other techniques.  We therefore decided to use the L-M results and assess the uncertainties using the ``overdispersion factor'' described in \S3.2.

\subsection{Pixel level decorrelation (PLD)}
The pixel level decorrelation technique \citep{2015ApJ...805..132D} is a parametric method that expresses the correlated noise in terms of a Taylor expansion sum of individual pixel values, instead of a function of centroid position.  The Taylor expansion partial derivatives become linear coefficients multiplying the (normalized) pixel values, a function that can be fit and removed.  As with ICA, using the individual pixel values avoids the flux/centroid degeneracy inherent in most decorrelation methods.  

The PLD reduction of the Data Challenge observations, by D. Deming, used 2D Gaussian centroiding \citep{2010ApJ...721.1861A} only to determine where to place the circular aperture, but not as a decorrelation variable.  An eclipse function was fit to the photometry simultaneously with the pixel coefficients and a quadratic phase curve \citep[see][Equation 4]{2015ApJ...805..132D}.  Due to time limitations, a full MCMC analysis of uncertainties was not possible, so error bars were estimated using the slope of the standard deviation vs. bin size relationship for the residuals \citep[as described in][]{2015ApJ...805..132D} to extrapolate to bins the width of the eclipse duration.  

\subsection{Segmented Polynomial, K2 Pipeline [SP(K2)]}
The segmented polynomial algorithm was originally developed for use with K2 data \citep{Buzasi:2015voa}, where detrending is normally required due to the presence of spacecraft pointing resets, and other less significant sources of correlated noise. The approach is reminiscent of polynomial surface fitting as used on \Spitzer\ data, but with some differences.  Detrending is carried out in two stages.  In the first stage, a third-order polynomial is fit to the flux vs.\,($x,y$) centroid for the entire time series and removed.  This process is repeated, with successive third order polynomial fits being applied to each set of residuals, until there is $< 1\%$ further reduction in the high-frequency noise standard deviation. In the second stage, the resulting time series is divided into segments, each of which is iteratively decorrelated using polynomial fitting. This segmented detrending is repeated for 10 different segment lengths between 0.04 and 0.125 of the total time series length, and the final time series is the result of applying a median filter to the 10 results.

SP(K2) detrending was applied to the \Spitzer\ Data Challenge measurements by D. Buzasi.  Due to time limitations a simple box function was used to fit the eclipse profile, and the uncertainties reported are formal fit errors.  

No attempt was made to tune SP(K2) for \Spitzer\ data. Future analysis might benefit from adjustment of the segmentation and fitting strategy.  For K2, the segmentation is partially necessary to accommodate unpredictable discontinuous jumps in source position when the pointing is reset, but this is not as much of an issue for \Spitzer\ staring mode observations (except for observations longer than 12 hours, for which there will be predictable pointing resets).  Furthermore, IRAC's intra-pixel gain variations are much larger than those of K2---the gain of IRAC varies by $\sim 8\%$ in Ch 1 (3.6\mum) and $\sim 4\%$ in Ch 2 (4.5\mum) across a pixel, whereas the K2 effect is only about 2\%.  \Spitzer\ data might be more amenable to spatial, rather than temporal, segmentation of the data in stage 2.\\\\



\bibliographystyle{aasjournal}
\bibliography{PapersLibrary}
\clearpage








\end{document}